\documentclass[prd,aps,showpacs,superscriptaddress,nofootinbib,amsmath,amssymb]{revtex4}
\usepackage{graphics}
\usepackage{graphicx}
\usepackage{dcolumn}
\usepackage{bm}
\usepackage{lipsum}
\usepackage{multirow}
\usepackage{adjustbox}
\usepackage{textcomp}
\graphicspath{{./eps/}}
\begin{document}
\title{Polarization observables and T-noninvariance in the weak charged current induced electron proton scattering}
\author{A. \surname{Fatima}}
\affiliation{Department of Physics, Aligarh Muslim University, Aligarh-202002, India}
\author{M. Sajjad \surname{Athar}}
\email{sajathar@gmail.com}
\affiliation{Department of Physics, Aligarh Muslim University, Aligarh-202002, India}
\author{S. K. \surname{Singh}}
\affiliation{Department of Physics, Aligarh Muslim University, Aligarh-202002, India}

\def\a              {\alpha}
\def\b              {\beta}
\def\m              {\mu}
\def\n              {\nu}
\def\ti             {\tilde}
\def\e              {\varepsilon}
\def\r              {\rho}
\def\s              {\sigma}
\def\g              {\gamma}
\def\d              {\delta}
\def\t              {\tau}

\begin{abstract}
In this work, we have studied the total scattering cross section ($\sigma$), differential scattering cross section ($d\sigma/dQ^2$) 
as well as the longitudinal ($P_L(E_e,Q^2)$), perpendicular ($P_P(E_e,Q^2)$), and transverse ($P_T(E_e,Q^2)$) components of the 
polarization of the final hadron ($n$, $ \Lambda$ and $\Sigma^0$) produced in the electron proton scattering induced by the weak 
charged current. We have not assumed T-invariance which allows the transverse component of the hadron polarization perpendicular to 
the production plane to be non-zero. The numerical results are presented for all the above observables and their dependence on the 
axial vector form factor and the weak electric form factor are discussed. The present study enables the determination of the axial 
vector nucleon-hyperon transition form factors at high $ Q^2$ in the strangeness sector which can provide test of the symmetries of 
the weak hadronic currents like T-invariance and SU(3) symmetry while assuming the hypothesis of conserved vector current and partial 
conservation of axial vector current. 
\end{abstract}

\maketitle
\section{Introduction}
The study of the nucleon polarizations in the elastic scattering of electron from the (un)polarized proton targets has made important 
contributions to our present understanding of the electromagnetic form factors of the nucleons in the region of high 
$Q^2$~\cite{Punjabi:2015bba, Pacetti:2015iqa}. In the case of the neutrino-nucleon scattering processes induced by the weak charged 
currents, various suggestions have been made in the past by many authors that the polarization measurements of the leptons and hadrons 
produced in the final state along with the measurements of the differential cross sections can provide a determination of all the 
weak nucleon form factors~\cite{Lee:1962jm, Block:1965zol, Block:NAL, Adler:1963, Ketley, Berman:1964zza, Fujii2, Fujii1, 
Cannata:1970br, Doncel:1971qb, Okamura:1971pn, DeRujula:1970ek, Pais:1971er, LlewellynSmith:1971uhs, Javannic:NAL, Marshak, 
Block:1964gj, Cabibbo:1965zza}. A precise determination of these form factors in the (anti)neutrino-nucleon scattering would 
facilitate the study of various symmetry properties and conservation laws associated with the weak hadronic currents in the 
$\Delta S = 0$ and $\Delta S = 1$ sectors.

While discussing the importance of the polarization measurements in the (anti)neutrino-nucleon scattering in the development of the 
weak interaction theory, some of the earlier works~\cite{Block:1965zol, Block:NAL, Pais:1971er, LlewellynSmith:1971uhs, Marshak, 
Javannic:NAL} have emphasized the technical challenges and difficulties involved in making such polarization measurements of the 
final nucleon in reactions like $\bar{\nu}_{\mu} + p \longrightarrow \mu^+ + n$ and $\nu_{\mu} +n \longrightarrow \mu^- + p$ with 
the experimental facilities then available at the CERN and other laboratories involved in doing (anti)neutrino scattering experiments. 
In this context, it was suggested that the observation of the final hadron polarizations may be feasible if a hyperon ($\Lambda$ 
or $\Sigma^0$) is produced in the quasielastic antineutrino-nucleon reactions like $\bar{\nu}_{\mu} +  p \longrightarrow \mu^+ + 
\Lambda (\Sigma^0)$~\cite{Adler:1963, Pais:1971er, LlewellynSmith:1971uhs, Marshak, Okamura:1971pn, Egardt}. This is because the 
$\Lambda$ or $\Sigma^0$ produced in these reactions decay into pions and the asymmetry in the angular distribution of the pions with 
respect to a given direction depends upon the polarization of $\Lambda$ or $\Sigma^0$ in that direction. Therefore, an observation of 
the asymmetry in the angular distribution of the pions determines directly the polarization of $\Lambda (\Sigma^0)$. Indeed, an 
experiment to study the quasielastic hyperon production performed at CERN with the SPS antineutrino beam has reported the results on 
the longitudinal, perpendicular and transverse polarizations of $\Lambda$~\cite{Erriquez:1978pg} along with the cross section 
measurements while the other experiments performed at CERN, BNL, FNAL and SKAT have published their results only on the cross section 
measurements~\cite{Erriquez:1977tr, Eichten:1972bb, Fanourakis:1980si, Ammosov:1986jn, Ammosov:1986xv, Brunner:1989kw}. However, 
the experimental uncertainties quoted in the values of the form factors determined in the axial vector sector were quite large due 
to the poor statistics of the observed $\Lambda$ events.

In recent years, there has been considerable development in the field of experimental neutrino physics with the availability of 
high intensity neutrino beams and technical advances in designing the large volume detectors to study the physics of neutrino 
oscillations. The specific experimental programs dedicated to make such studies also plan to perform (anti)neutrino-nucleon cross 
section measurements in the near detector for various reaction channels in the scattering of (anti)neutrinos from the nuclear 
targets~\cite{Longhin:2017tfq, Furmanski:2017plq, Abi:2017aow, Varanini:2017pvw}. These experiments may also be able to study the
hyperon production with better statistics and perform the polarization measurements for the final leptons and hadrons. In this 
context, there have been many theoretical calculations of the polarization observables of the final leptons and hadrons in 
the (anti)neutrino-nucleon scattering processes~\cite{Graczyk:2017rti, Akbar:2016awk, Kuzmin:2003ji, Graczyk:2004uy, Hagiwara:2004gs, 
Graczyk:2004vg, Bilenky:2013fra, Bilenky:2013iua, Valverde:2006yi,Kuzmin:2004ke}.

As shown in some of these calculations the independent determination of in-the-plane, {\it i.e.,} longitudinal and 
perpendicular components of the polarization and their $Q^2$ dependence may be helpful in resolving the present tension 
between the values of the dipole mass in the parameterization of the axial vector form factor as determined presently 
from the medium energy and high energy (anti)neutrino-nucleus scattering 
experiments~\cite{Katori:2016yel, Morfin:2012kn, Formaggio:2013kya}. On the other hand, the measurement of the 
transverse component of the polarization, which is in the direction perpendicular to the plane of reaction, gives 
information about the weak electric form factor which is forbidden by the T-invariance. Such a measurement will provide 
an opportunity to study the physics of T-noninvariance and test the theoretical models proposed for T-noninvariance in 
neutrino interactions~\cite{Cabibbo:1964zza, Glashow:1965zz}.

However, the interpretation of the neutrino-nucleus scattering results on the differential cross sections and the angular distribution 
as well as the energy distribution of the final leptons and hadrons in the quasielastic and inelastic neutrino reactions has been 
beset with the systematic uncertainties arising due to the use of the continuous neutrino energy beams and the heavy nuclear targets 
in these experiments~\cite{Katori:2016yel, Morfin:2012kn, Formaggio:2013kya}. Therefore, the interpretation of the polarization 
measurements, if made in future, in these experiments may also be subject to these uncertainties.

In view of the systematic uncertainties associated with the interpretation of the neutrino scattering experiments, some suggestions 
have been made in the past to study the weak interactions with the high intensity monoenergetic electron beams using hydrogen 
targets~\cite{Fearing:1969nr, Hwang:1988fp}. The feasibility of doing such experiments has been recently studied in the
literature~\cite{Mintz:2004eu, Mintz:2002cj, Mintz:2001jc, Akbar:2017qsf} with the availability of high intensity monoenergetic 
electron beams at MAMI and JLab electron accelerator facilities~\cite{jlab, mainz}. The advantage of studying the hyperon production 
at lower electron energies is that there exists no background of $\Lambda(\Sigma^0)$ below the threshold of the associated particle 
production, $e^- + p \longrightarrow e^- + \Lambda(\Sigma^0) + K^+ $, which is around 910~MeV~(1050 MeV). We have recently studied 
theoretically the cross sections and polarizations of hyperon produced in the reaction 
$e^- + p \longrightarrow \Lambda(\Sigma^0) + \nu_e$, assuming T-invariance~\cite{Akbar:2017qsf}. 

In this paper, we present a full calculation of the cross sections and all the polarization observables for the final hadron 
produced in the reactions $e^- + p \longrightarrow n + \nu_e$ and $e^- + p \longrightarrow \Lambda (\Sigma^0) + \nu_e$ without 
assuming T-invariance, with the aim to study the physics of T-noninvariance in weak interactions. We assume a model of 
T-noninvariance in the weak hadronic currents due to Cabibbo~\cite{Cabibbo:1964zza}.

In section~\ref{formalism}, we briefly describe the formalism for calculating the differential and the total scattering cross 
sections, and all the components of the final hadron polarization in the electron proton scattering induced by the weak charged 
currents in $\Delta S=0$ and $\Delta S=1$ sectors. We calculate the matrix elements of the weak hadronic currents using the 
assumption of the Conserved Vector Current (CVC), the Partial Conservation of Axial Vector Current (PCAC), and charge symmetry in the 
$\Delta S = 0$ sector and extending their validity to the $\Delta S=1$ sector using SU(3) symmetry in the presence of 
T-noninvariance~\cite{Pais:1971er,LlewellynSmith:1971uhs, Marshak, Cabibbo:2003cu, Block:1964gj, Cabibbo:1965zza}. In 
section~\ref{results}, we present and discuss the numerical results. The conclusions are summarized in section~\ref{conclusion}.

\section{Weak $\bm{e-p}$ scattering induced by the charged current}\label{formalism}
\subsection{Matrix element and cross section}
  \begin{figure}
 \begin{center}
    \includegraphics[height=3cm,width=6cm]{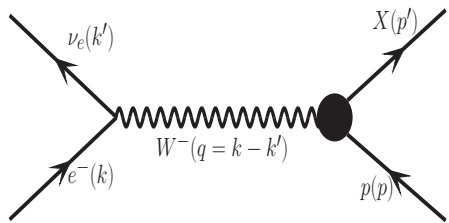}
  \caption{Feynman diagram  for the process $ e^-(k) + p(p) \rightarrow \nu_e(k^\prime) + X(p^{\prime})$; $X = n, \Lambda, \Sigma^0$. 
  The quantities in the bracket represent 
  four momenta of the corresponding particles.}\label{fyn_hyp}
   \end{center}
 \end{figure}
The transition matrix element for the processes,
\begin{eqnarray}
 e^- (k) + p (p) &\longrightarrow& \nu_e (k^\prime) + n (p^{\prime}), ~~~~~~~~~~~~~~~~~~~~~~~~~~~ (\Delta S=0) \label{nuc-rec}, \\
 e^- (k) + p (p) &\longrightarrow& \nu_e (k^\prime) + Y (p^{\prime}), ~~~~~ Y = \Lambda,\Sigma^0,~~~~~~~ (\Delta S=1), \label{hyp-rec}
\end{eqnarray}
 presented in Fig.~(\ref{fyn_hyp}), may be written as
 \begin{eqnarray}\label{matrix}
  {\cal{M}} = \frac{G_F}{\sqrt{2}} a~ l^\mu {{J}}_\mu,
 \end{eqnarray}
 where the quantities in the brackets (Eqs. (\ref{nuc-rec}) and (\ref{hyp-rec})) represent the four momenta of the corresponding 
 particles, $G_F$ is the Fermi coupling constant, $a = \cos \theta_c$ for $\Delta S = 0$ processes, $a = \sin \theta_c$ for 
 $\Delta S = 1$ processes, and $\theta_c$ is the Cabibbo mixing angle. The leptonic current $l^\mu$ is given by 
 
\begin{equation}
 l^\mu = \bar{u} (k^\prime) \gamma^\mu (1-\gamma_5) u (k),
\end{equation}
and the hadronic current ${J}_\mu$ is expressed in terms of the vector and the axial vector currents as~\cite{LlewellynSmith:1971uhs}:
\begin{equation}
 {{J}}_\mu =  \bar{u} (p^\prime) (V_\mu - A_\mu) u (p)
\end{equation}
with
\begin{eqnarray}\label{vx}
 V_\mu &=& \gamma_\a f_1^{pX}(Q^2)+i\sigma_{\a\b} \frac{q^\b}{M+M^\prime} f_2^{pX}(Q^2)
  + \frac{2 ~q_\a}{M+M^\prime} f_3^{pX}(Q^2),
  \end{eqnarray}
and 
\begin{eqnarray}\label{vy}
  A_\mu &=&  \gamma_\a \gamma_5 g_1^{pX}(Q^2) + i \sigma_{\a\b}\gamma_5 \frac{q^\b}{M+M^\prime} g_2^{pX}(Q^2)  
   + \frac{2 ~q_\a} {M+M^\prime} g_3^{pX}(Q^2) \gamma_5 , \text{}
\end{eqnarray}
where X stands for a nucleon $N (=n)$ or a hyperon $Y(=\Lambda, \Sigma^0)$, $M$ and $M^\prime$ are the masses of the initial and 
final hadrons with $M^\prime = M$ for the nucleon and $M^\prime = M_\Lambda$ or $M_\Sigma$ for the hyperons. 
$q_\mu (= k_\mu - k_\mu^\prime = p_\mu^\prime -p_\mu)$ is the four momentum transfer with $Q^2 = - q^2, Q^2 >0$. $f_{1}^{pX}(Q^2)$, 
$f_{2}^{pX}(Q^2)$ and $f_{3}^{pX}(Q^2)$ are the vector, weak magnetic and scalar form factors and $g_{1}^{pX}(Q^2)$, 
$g_{2}^{pX}(Q^2)$ and $g_{3}^{pX}(Q^2)$ are the axial, weak electric and pseudoscalar form factors, respectively and are discussed 
below in detail.

 \subsubsection{Form Factors}\label{sec:form}
 The six form factors $f_i^{pX} (Q^2)$ and $g_i^{pX} (Q^2) ~ (i=1-3)$ are determined using the following assumptions about the 
 symmetry properties of the vector and the axial vector currents in the weak interactions~\cite{Pais:1971er,LlewellynSmith:1971uhs, 
 Marshak, Cabibbo:2003cu, Block:1964gj, Cabibbo:1965zza}:
 
 \begin{itemize}
  \item [(i)] The requirement of T-invariance implies that all the vector $f_i^{pX} (Q^2) (i=1-3)$ and the axial vector 
  $g_i^{pX} (Q^2) (i=1-3)$ form factors are real apart from an arbitrary overall phase factor which is generally taken to be real.
  
  \item[(ii)] The assumption of $\Delta S=0$ weak hadronic currents belonging to the isotriplet implying charge symmetry requires that 
  $f_1^{pX} (Q^2)$, $f_2^{pX} (Q^2)$, $g_1^{pX} (Q^2)$ and $g_3^{pX} (Q^2)$ be real, while $f_3^{pX} (Q^2)$ and $g_2^{pX} (Q^2)$ be 
  imaginary.
  
  \item[(iii)] The requirement of the combined symmetry of T-invariance and C-invariance with charge symmetry (or according to the 
  G-transformation due to Weinberg~\cite{Weinberg:1958ut} or CP-transformation due to Cabibbo~\cite{Cabibbo:1964zza}) requires that 
  the form factors $f_3^{pX} (Q^2) = 0$ and $g_2^{pX} (Q^2) = 0$ which correspond to the second class currents or the irregular 
  currents. This classification is strictly true for the isotriplet of $\Delta S = 0$ currents but can be extended to the octet of 
  vector and axial vector currents in the limit of the exact SU(3) symmetry and is, therefore, applicable to the $\Delta S = 1$ 
  currents as well~\cite{Cabibbo:2003cu, Cabibbo:1965zza}.  
  
  \item[(iv)] The hypothesis of CVC implies that $f_3^{pX} (Q^2) = 0$.
  
  \item[(v)] The hypothesis of PCAC implies that $g_3^{pX} (Q^2)$ can be related to $g_1^{pX} (Q^2)$ in terms of the pion pole 
  (in the case of $\Delta S = 0$ current) or the kaon pole (in the case of $\Delta S = 1$ current)~\cite{Goldberger:1958vp, 
  Marshak, Nambu:1960xd}. However, we take $ g_3^{pX} (Q^2) = 0$ in the numerical calculations as its contribution is proportional 
  to the lepton mass and is almost negligible in the case of the reactions involving electron and $\nu_e$.  
  
  \item[(vi)] It should be noted that while the existence of a purely imaginary $g_2^{pX} (Q^2)$ implies a violation of T-invariance, 
  a purely real $g_2^{pX} (Q^2)$ is consistent with T-invariance while giving G-violation along with violation of charge symmetry as 
  in the model of Ref.~\cite{Ohtsubo}. A complex value of $g_2^{pX} (Q^2)$, therefore, implies the violation of the G-invariance and 
  charge symmetry which also leads to the violation of T-invariance as in the model proposed by Glashow~\cite{Glashow:1965zz}. While the 
  implications of the G-invariance are strictly valid for $\Delta S=0$ currents with SU(2) symmetry, they are not applicable to the octet 
  currents due to SU(3) breaking, which may induce G-violating effects. Indeed, while the experimental limits on the existence of 
  $g_2^{pn} (Q^2)$ are quite small in the $\Delta S = 0$ sector~\cite{Ahrens:1988rr}, they are not so small in the case of 
  $g_2^{pX} (Q^2)$ corresponding to the $\Delta S=1$ semileptonic hyperon decays~\cite{Cabibbo:2003cu, Holstein}.
  
 \end{itemize}
 
 We now elaborate the implications of assuming SU(3) symmetry in determining the form factors $f_i^{pX} (Q^2) (i=1,2)$ and 
 $g_i^{pX} (Q^2)(i=1,2)$ in terms of the nucleon form factors in the presence of T-noninvariance. The assumption of the SU(3) 
 symmetry of the weak hadronic currents implies that the vector and the axial vector currents transform as an octet under the 
 SU(3) group of transformations.

\begin{enumerate} 

\item[a)] Since the initial and final baryons also belong to the octet representation, therefore, each form factor 
$f^{pX}_i(Q^2)$ and $g^{pX}_i(Q^2)$ occurring in the matrix element of the hadronic current is written in terms of the two functions 
$D(Q^2)$ and $F(Q^2)$ corresponding to the symmetric octet($8^{S}$) and antisymmetric octet($8^{A}$) couplings of octets of the
vector and the axial vector currents. Specifically, we write
\begin{eqnarray}
 f^{pX}_i(Q^2) &=& a F_{i}^{V}(Q^2) + b D_{i}^{V}(Q^2)\label{coef1},\\
 g^{pX}_i(Q^2) &=& a F_{i}^{A}(Q^2) + b D_{i}^{A}(Q^2)\label{coef2},
\end{eqnarray}
where $a$ and $b$ are SU(3) Clebsch-Gordan coefficients given in Table-\ref{tab1}. $F_i^{V,A}(Q^2)$ and $D_i^{V,A}(Q^2);~(i=1,2)$, 
are the couplings corresponding to the antisymmetric and symmetric couplings of the two octets.

\item[b)] The two vector form factors \textit{viz.} $f^{pX}_1(Q^2)$ and $f^{pX}_2(Q^2)$ are determined in terms of the 
electromagnetic form factors of the nucleon, \textit{i.e.} $f_{1}^{N}(Q^{2})$ and $f_{2}^{N}(Q^{2}),~ N=(p,n)$. This is done by 
taking the matrix element of the electromagnetic current operator between the nucleon states and determining $F_{i}^{V}(Q^2)$ and 
$D_{i}^{V}(Q^2)$ in terms of $f^{N}_{i}(Q^2);~(i = 1, 2)$. The functions $F_{i}^{V}(Q^2)$ and $D_{i}^{V}(Q^2)$ are, thus, 
expressed in terms of the nucleon form factors $f_{1}^{p,n}(Q^{2})$ and $f_{2}^{p,n}(Q^{2})$ as 
\begin{eqnarray}\label{eq:fiv_div}
 F_i^V(Q^2) &=& f_i^p(Q^2) + \frac12 f_i^n (Q^2),  \\ 
 D_i^V(Q^2) &=& - \frac32 f_i^n (Q^2). 
\end{eqnarray}
 $f_{1}^{N}(Q^{2})$ and $f_{2}^{N}(Q^{2})$ are expressed in terms of the experimentally determined Sachs' electric $G_E^{p,n} (Q^2)$ 
 and magnetic $G_M^{p,n}(Q^2)$ form factors for which various parameterizations are available in the literature but we have used 
 the parameterization given by Bradford et al.~\cite{Bradford:2006yz} in the numerical calculations.
 
\begin{table}[h!]\centering
 \begin{tabular}{|c|c|c|c|c}\hline
 &~$p \to n $& ~$p \to \Lambda$~ &  ~$ p \to \Sigma^{0}$~ \\ \hline \hline 
~~$a$&  ~~~~~1~~~~~~& ~~~$-\sqrt{\frac{3}{2}}$~~~ & ~~~~$-\frac{1}{\sqrt{2}}$~~~~ \\ \hline 
 ~~$b$& ~~~~~1~~~~~~& ~~~$-\sqrt{\frac{1}{6}}$~~~ & ~~~~$\frac{1}{\sqrt{2}}$~~\\ \hline 
 
 \end{tabular}
\caption{Values of the coefficients $a$ and $b$ given in Eqs.~(\ref{coef1}) and (\ref{coef2}).}
\label{tab1}
\end{table}

\item[c)]  The axial vector form factors $g^{pX}_{1}(Q^{2})$ and $g^{pX}_{2}(Q^{2})$ are determined from Eq.~(\ref{coef2}).
$g^{pX}_{1,2}(Q^2)$ are written in terms of the two functions $F_{1,2}^A(Q^2)$ and $D_{1,2}^A(Q^2)$. Using Table-\ref{tab1} 
for the coefficients $a$ and $b$, we find 
\begin{eqnarray}
   g_{1,2}^{pn}(Q^2)&=& F_{1,2}^A(Q^2)+D_{1,2}^A(Q^2), \\
  g_{1,2}^{p\Lambda}(Q^2)&=&\sqrt{\frac16}\left(3F^A_{1,2}(Q^2)+D^A_{1,2}(Q^2)\right), \\
   g_{1,2}^{p\Sigma^0}(Q^2)&=& \sqrt{\frac12} \left[ D^A_{1,2}(Q^2)-F^A_{1,2}(Q^2) \right].
\end{eqnarray}\label{Eq:xdep}
$g_{1,2}^{p\Lambda}(Q^2)$ and $g_{1,2}^{p\Sigma^0}(Q^2)$ are rewritten in terms of $g_{1,2}^{pn}(Q^2)$ and $x_{1,2}(Q^2)$, 
with $x_{1,2}(Q^2)$ defined as 
\begin{equation}
x_{1,2}(Q^2)=\frac{F^A_{1,2}(Q^2)}{F^A_{1,2}(Q^2)+D^A_{1,2}(Q^2)}.
\end{equation}

We further assume that $F^A_{1,2}(Q^2)$ and $D^A_{1,2}(Q^2)$ have the same $Q^2$ dependence, such that $x_{1,2}(Q^2)$ become 
constant given by $x_{1,2}(Q^2)=x_{1,2}=\frac{F^A_{1,2}(0)}{F^A_{1,2}(0)+D_{1,2}^A(0)}$. 

The various form factors $f_{i=1,2}^{pX} (Q^2)$ and $g_{i=1,2}^{pX} (Q^2)$ are, thus, written in terms of $f_{i=1,2}^{p,n} (Q^2)$ 
and $g_{i=1,2}^{pn} (Q^2)$, which are given in 
Table-\ref{tab:formfac}.

\begin{table}[h!]
 \begin{center}
\begin{adjustbox}{max width=\textwidth} 
\begin{tabular}{|c|c|c|c|}  \hline 
&$e^- p \rightarrow \nu_e X (=n)$&$e^- p \rightarrow \nu_e X( =\Lambda)$&$e^- p \rightarrow \nu_e X=(\Sigma^0)$\\ \hline  \hline     
 $f_1^{pX}(Q^2)$& $f_1^p(Q^2) - f_1^n(Q^2)$&$ -\sqrt{\frac{3}{2}}~f_1^p(Q^2)$&$-\frac{1}{\sqrt2}\left[f_1^p(Q^2) + 2 f_1^n(Q^2) 
 \right]$ \\ \hline
$f_2^{pX}(Q^2)$& $f_2^p(Q^2) - f_2^n(Q^2)$&$-\sqrt{\frac{3}{2}}~f_2^p(Q^2)$&$-\frac{1}{\sqrt2}\left[f_2^p(Q^2) + 2 f_2^n(Q^2) 
\right]$\\ \hline
$g_1^{pX}(Q^2)$& $g_1^{pn}(Q^2)$&$-\frac{1}{\sqrt{6}}(1+2x_1) g_1^{pn}(Q^2)$&$\frac{1}{\sqrt2}(1-2x_1)g_1^{pn}(Q^2)$ \\ \hline 
 $g_2^{pX}(Q^2)$& 
 $~g_2^{pn}(Q^2)$&$- \frac{1}{\sqrt{6}}(1+2x_2) g_2^{pn}(Q^2)$&$ \frac{1}{\sqrt2}(1-2x_2)g_2^{pn}(Q^2)$\\ \hline 
  \end{tabular}
  \end{adjustbox}
\end{center}
\caption{Vector and axial vector from factors $ f_i^{pX} (Q^2)$ and $g_i^{pX} (Q^2)~ (i=1,2)$, $f_3^{pX}(Q^2)$ and $g_3^{pX} (Q^2)$ 
are taken to be zero and negligible respectively (see text) for the $e^-(k) + p(p)\rightarrow \nu_e(k^\prime) + X(p^\prime)$ 
processes, where $X=n, \Lambda^0, \Sigma^0$.}
 \label{tab:formfac}
\end{table}

\item [d)] For the axial vector form factor $g_{1}^{np}(Q^2)$, a dipole parameterization has been used:
\begin{eqnarray}\label{g1}
 g_{1}^{pn}(Q^2)=g_{A}(0)\left(1+\frac{Q^2}{M_{A}^2}\right)^{-2},
\end{eqnarray}
where $M_A$ is the axial dipole mass and $g_A(0)$ is the axial charge. For the numerical calculations, we have used the world average 
value of $M_A=1.026$ GeV~\cite{Bernard:2001rs} unless stated. $g_A(0)$ and $x_1$ are taken to be 1.2723 and 0.364, respectively, as 
determined from the experimental data on the $\beta-$decay of neutron and the semileptonic decay of hyperons~\cite{Cabibbo:2003cu}. 

\item [e)] The weak electric form factor $g_2^{pn} (Q^2)$ is taken to be of dipole form, i.e., 
\begin{eqnarray}\label{g2} 
 g_{2}^{pn}(Q^2)=g_{2}^{pn}(0)\left(1+\frac{Q^2}{M_{2}^2}\right)^{-2}.
 \end{eqnarray}
There is limited experimental information about $g_2^{pn} (Q^2)$ which is obtained from the analysis of the weak processes while searching for 
 the G-noninvariance assuming T-invariance~\cite{Holstein,Cabibbo:2003cu}. As a consequence of T-invariance this information is useful in 
estimating the range of the real values of $g_2^{pn} (Q^2)$. This is not relevant for our present purpose as we are concerned with the 
T-violating effects and used Cabibbo's model where $g_2^{pn}(Q^2)$ is purely imaginary~\cite{Cabibbo:1964zza}.

Our present knowledge about the $Im~ g_2^{pX} (Q^2); (X=n,\Lambda,\Sigma^0)$ is almost non-existent. The analysis of T-violating polarization 
observables in the neutrino scattering in the CERN experiment~\cite{Erriquez:1978pg} quotes a value of the T-violating polarization observable 
consistent with zero but no limits on $Im~g_2^{p \Lambda} (Q^2)$ are reported. Similarly, an older analysis of the T-violating spin 
correlations of electrons in the beta decay of polarized $\Lambda$ implies a large value of $Im~g_2^{p \Lambda} (Q^2)$ without giving a 
quantitative estimate~\cite{Oehme:1971rd}. While there are no theoretical model studies of $Im~g_2^{p X} (Q^2)$, the 
older calculations of T-violating effects in weak processes have phenomenologically used the values of $Im~g_2^{p X} (Q^2)$ in a large range 
of $1<Im~g_2^{pn}(0)<10$~\cite{Fearing:1969nr} while $Im~g_2^{pn} (0) = f_2^{pn}(0)$~\cite{Berman:1964zza} and 
$Im~g_2^{p \Lambda}(0) = 1.92$~\cite{Fujii1} have also been used.

In view of the above discussion, we have used a value of $Im~g_2^{pn}(0)$ in the range $1<Im~g_2^{pn} (0)<3$ and $x_2 = x_1(=0.364)$. In case 
of two parameter description of $g_2^{pX} (0)$ where $g_2^{pn} (0)$ and $x_2 (\neq x_1)$ can be treated as free parameters, we have also 
varied the value of $x_2$ by $20\%$ around $x_1(=0.364)$. We have also studied the implications of considering $Im~g_2^{pn}(0)$ to be positive 
or negative.

In this work our main aim is to study the T-violating effects, therefore, we do not consider a complex value of $g_2^{pX} (Q^2)$ as a real 
component of $g_2^{pX} (Q^2)$ will give contributions only to the T-invariant observables like rates, angular and energy distributions as well 
as the longitudinal and perpendicular components of the hadron polarization in the reaction plane.

\end{enumerate}
\subsubsection{Cross section}
\begin{figure}
 \begin{center}
   \hspace{-1cm}
   \includegraphics[height=6cm,width=7cm]{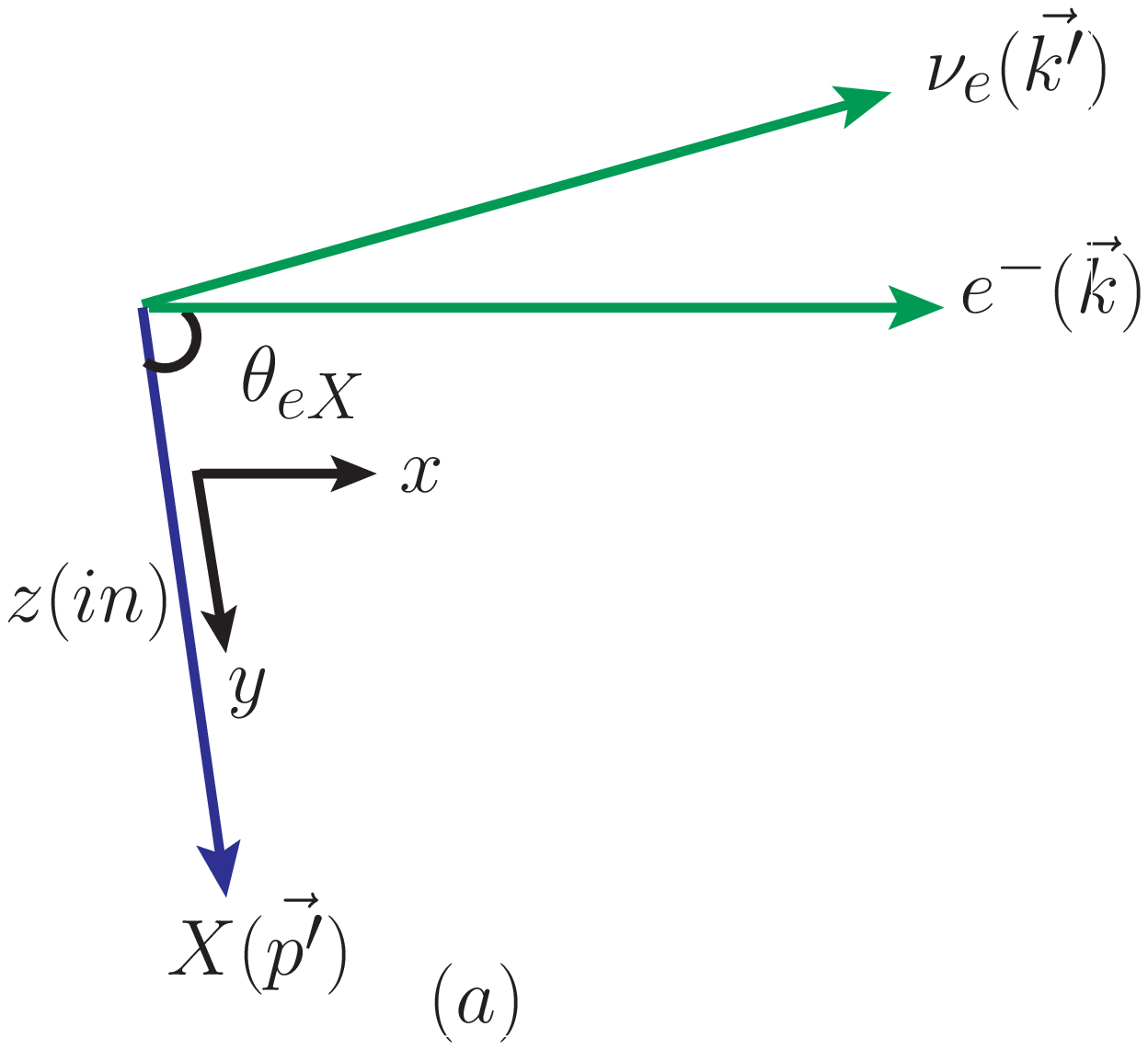}
    \hspace{1cm}
        \includegraphics[height=6cm,width=7cm]{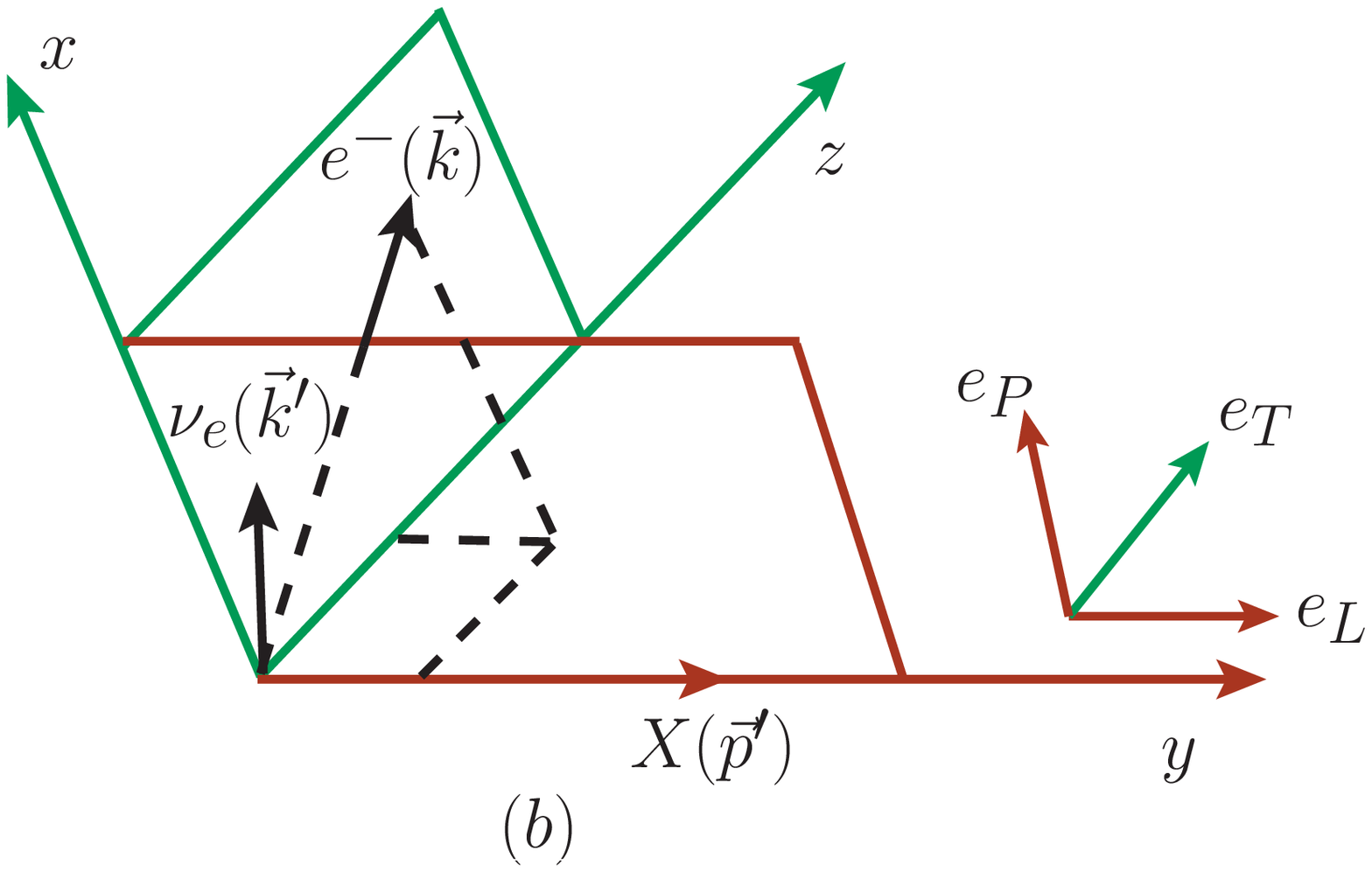}
  \caption{(a)~Diagrammatic representation of the process $ e^-(\vec{k}) + p(\vec{p}=0) \rightarrow \nu_e(\vec{k^\prime}) + 
  X(\vec{p}^{~\prime})$; $X =n, \Lambda, \Sigma^0$. (b)~The longitudinal, perpendicular and transverse directions with respect to 
  the momentum of the final hadron.}\label{TRI}
   \end{center}
 \end{figure}

The general expression of the differential cross section corresponding to the processes (\ref{nuc-rec}) and (\ref{hyp-rec}), depicted
in Fig.~\ref{TRI}(a) in the rest frame of the initial proton, is written as:
 \begin{eqnarray}
 \label{crosv.eq}
 d\sigma&=&\frac{1}{(2\pi)^2}\frac{1}{4E_e M}\delta^4(k+p-k^\prime-p^\prime) \frac{d^3k^\prime}{2E_{k^\prime}}  
 \frac{d^3p^\prime}{2E_{p^\prime}} \overline{\sum} \sum |{\cal{M}}|^2,
 \end{eqnarray}
 where $E_e$ is the electron energy and the transition matrix element squared is expressed as:
\begin{equation}\label{matrix}
  \overline{\sum} \sum |{\cal{M}}|^2 = \frac{G_F^2 a^2}{2} \cal{L}_{\alpha \beta} \cal{J}^{\alpha \beta}.
\end{equation}
The leptonic and the hadronic tensors are given by
\begin{eqnarray}
\cal{L}^{\alpha \beta} &=& \frac{1}{2}\mathrm{Tr}\left[\gamma^{\alpha}(1-\gamma_{5})(k\!\!\!/+m_e)
\gamma^{\beta}(1-\gamma_{5})k'\!\!\!\!\!/~\right], \\ \label{L}
\cal{J}_{\alpha \beta} &=& \frac{1}{2} \mathrm{Tr}\left[\Lambda(\not{p^\prime}) J_{\alpha}
  \Lambda(\not{p}) \tilde{J}_{\beta} \right], \label{J}
\end{eqnarray} 
with $\Lambda(p)=(p\!\!\!/+M)$, $\Lambda(p^{\prime})=(p\!\!\!/^{\prime}+M^{\prime})$ and $\tilde{J}_{\beta} =\gamma^0 
J^{\dagger}_{\beta} \gamma^0$.

 Using the above definitions, the $Q^2$ distribution is written as
\begin{equation}\label{dsig}
 \frac{d\sigma}{dQ^2}=\frac{G_F^2 ~a^2}{16 \pi M^2 {E_e}^2} N(E_e,Q^2),
\end{equation}
where the expression of $N(E_e,Q^2)$ is given in the Appendix.

 \subsection{Polarization observables of the final hadrons and T-noninvariance} 
  Using the covariant density matrix formalism, the polarization 4-vector($\xi^\tau$) of the final hadron produced in the reactions 
  (\ref{nuc-rec}) and (\ref{hyp-rec}) is written as~\cite{Bilekny}:
 
\begin{eqnarray}\label{polar4}
\xi^{\tau}&=&\left( g^{\tau\sigma}-\frac{p'^{\tau}p'^{\sigma}}{{M^\prime}^2}\right) \frac{  {\cal L}^{\alpha \beta}  \mathrm{Tr}
\left[\gamma_{\sigma}\gamma_{5}\Lambda(p')J_{\alpha} \Lambda(p)\tilde{J}_{\beta} \right]}
{ {\cal L}^{\alpha \beta} \mathrm{Tr}\left[\Lambda(p')J_{\alpha} \Lambda(p)\tilde{J}_{\beta} \right]}.
\end{eqnarray}

One may write the polarization vector $\vec{\xi}$ in terms of the three orthogonal vectors $\hat{e}_{i}~(i=L,P,T)$, i.e.,
 \begin{equation}\label{polarLab}
\vec{\xi}=\xi_{L} \hat{e}_{L} + \xi_{P} \hat{e}_{P}+\xi_{T} \hat{e}_{T} ,
\end{equation}
where $\hat{e}_{L}$, $\hat{e}_{P}$ and $\hat{e}_{T}$ are chosen to be the set of orthogonal unit vectors corresponding to the 
longitudinal, perpendicular and transverse directions with respect to the momentum of the final hadron, shown in Fig~\ref{TRI}(b), 
and are written as
\begin{equation}\label{vectors}
\hat{ e}_{L}=\frac{\vec{ p}^{\, \prime}}{|\vec{ p}^{\, \prime}|},~~~~~
\hat{ e}_{P}=\hat{ e}_{L}\times \hat{ e}_T, ~~~~ 
\hat{e}_T=\frac{\vec{ p}^{\, \prime}\times \vec{ k}}{|\vec{ p}^{\, \prime}\times \vec{ k}|}.
 \end{equation}
The longitudinal, perpendicular and transverse components of the polarization vector $\vec{\xi}_{L,P,T} (Q^2)$ using Eqs. 
(\ref{polarLab}) and (\ref{vectors}) may be written as
\begin{equation}\label{PL}
 \xi_{L,P,T}(Q^2)=\vec{\xi} \cdot \hat{e}_{L,P,T}~.
\end{equation}
In the rest frame of the initial nucleon, the polarization vector $\vec{\xi}$ is expressed as

\begin{equation}\label{pol2}
 \vec{\xi} = A(E_e,Q^2)~ \vec{k} + B(E_e,Q^2)~ \vec{p}^{\, \prime} + C(E_e,Q^2)~  M (\vec{k} \times \vec{p}^{\,\prime})
\end{equation}
and is explicitly calculated using Eq.~(\ref{polar4}). The expressions for the coefficients $A(E_e,Q^2)$, $B(E_e,Q^2)$ and 
$C(E_e,Q^2)$ are given in the Appendix.

\begin{figure}
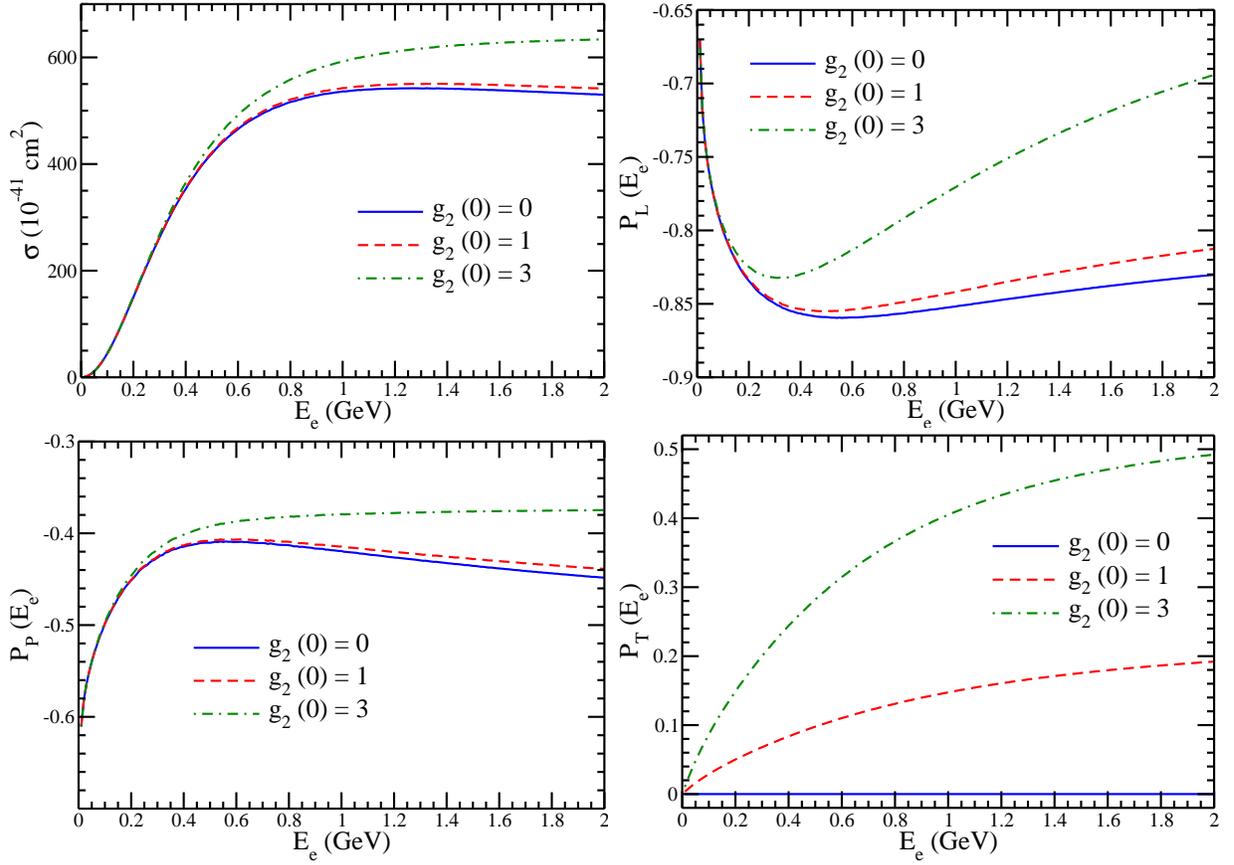

 \includegraphics[height=5.7cm,width=8cm]{total_sigma_g2_variation_neutron.eps}
 \includegraphics[height=5.7cm,width=8cm]{Pl_Ee_g2_variation_neutron.eps}  \\
 \includegraphics[height=5.7cm,width=8cm]{Pp_Ee_g2_variation_neutron.eps}
 \includegraphics[height=5.7cm,width=8cm]{Pt_Ee_g2_variation_neutron.eps}
\caption{$\sigma(E_e)~ vs. ~ E_e$ (upper left panel), $P_L (E_e) ~ vs. ~ E_e$ (upper right panel), $P_P (E_e) ~ vs. ~ E_e$ (lower 
left panel) and $P_T (E_e)~ vs.~ E_e$ (lower right panel) for the process ${e^- + p \rightarrow \nu_e + n}$ at $g_2 (0)$ = 
0~(solid line), 1~(dashed line), and 3~(dashed-dotted line) with $M_A = 1.026$ GeV.}\label{fig1}
\end{figure}

The longitudinal ($P_L(Q^2)$), perpendicular ($P_P(Q^2)$) and transverse ($P_T(Q^2)$) components of the polarization vector in the 
rest frame of the final hadron is then obtained by performing a Lorentz boost and is written as~\cite{Bilenky:2013fra}:

\begin{equation}\label{PlPp}
 P_L (Q^2) = \frac{M^\prime}{E^\prime} \xi_L (Q^2), ~~~~~~~ P_P (Q^2) = \xi_P (Q^2), ~~~~~~~ P_T (Q^2) = \xi_T (Q^2).
\end{equation}
The expressions for $P_L (Q^2)$, $P_P (Q^2)$ and $P_T (Q^2)$ are then obtained using Eqs.~(\ref{vectors}), (\ref{PL}) and 
(\ref{pol2}) in Eq.~(\ref{PlPp}) and are expressed as

\begin{eqnarray}
  P_L (Q^2) &=& \frac{M^\prime}{E^\prime} \frac{A(E_e,Q^2) \vec{k} \cdot \hat{p}^{\prime} + B (E_e,Q^2) |\vec{p}^{\,\prime}|}
  {N(E_e,Q^2)},
  \label{Pl} \\
 P_P (Q^2) &=& \frac{A(E_e,Q^2) [(\vec{k}.\hat{p}^{\prime})^2 - |\vec{k}|^2]}{N(E_e,Q^2) ~|\hat{p}^{\prime} \times \vec{k}|},
 \label{Pp} \\
  P_T (Q^2) &=& \frac{C(E_e,Q^2) M |\vec{p}^{\,\prime}|[(\vec{k}.\hat{p}^{\prime})^2 - |\vec{k}|^2]}{N(E_e,Q^2)~|\hat{p}^{\prime} 
  \times \vec{k}|}. \label{Pt}
\end{eqnarray}

\begin{figure}
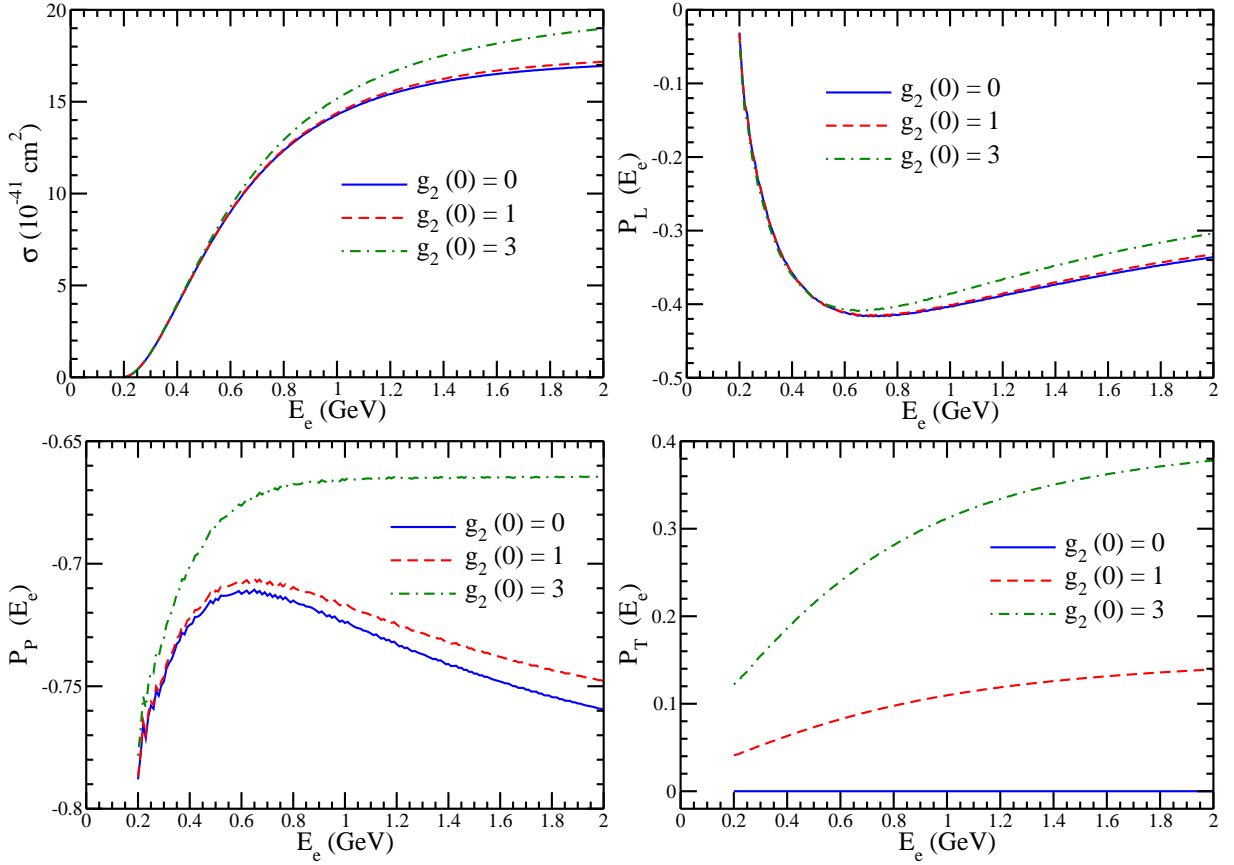

 \includegraphics[height=5.7cm,width=8cm]{total_sigma_g2_variation_lambda.eps}
 \includegraphics[height=5.7cm,width=8cm]{Pl_elep_g2_variation_lambda.eps}  \\
 \includegraphics[height=5.7cm,width=8cm]{Pp_elep_g2_variation_lambda.eps}
 \includegraphics[height=5.7cm,width=8cm]{Pt_elep_g2_variation_lambda.eps}
\caption{$\sigma(E_e)~ vs. ~ E_e$~(upper left panel), $P_L (E_e) ~ vs. ~ E_e$~(upper right panel), $P_P (E_e) ~ vs. ~ E_e$~(lower 
left panel) and $P_T (E_e)~ vs.~ E_e$~(lower right panel) for the process ${e^- + p \rightarrow \nu_e + \Lambda}$ at $g_2 (0)$ = 
0~(solid line), 1~(dashed line), and 3~(dashed-dotted line) with $M_A = 1.026$ GeV.}\label{fig2}
\end{figure}

If the T-invariance is assumed then all the vector and the axial vector form factors are real and the expression for $C(E_e,Q^2)$ 
vanishes which implies that the transverse component of polarization perpendicular to the production plane, $ P_T (Q^2)$  vanishes. 
If the T-invariance is violated then while $d\sigma/dQ^2$, $P_L(Q^2)$ and $P_P(Q^2)$ receive small corrections due to the T-violating 
form factor $g_2^{pX}(Q^2)$, the transverse component of the polarization $P_T (Q^2)$ becomes non-zero and could be significant. 
Thus, an experimental observation of $P_T (Q^2)$ can be used to study the physics of T-noninvariance assuming that the contributions 
to the T-violating effects, if any, induced by the electromagnetic corrections to the weak processes are small~\cite{DeRujula:1970ek, 
Fearing:1969nr}.

\begin{figure}
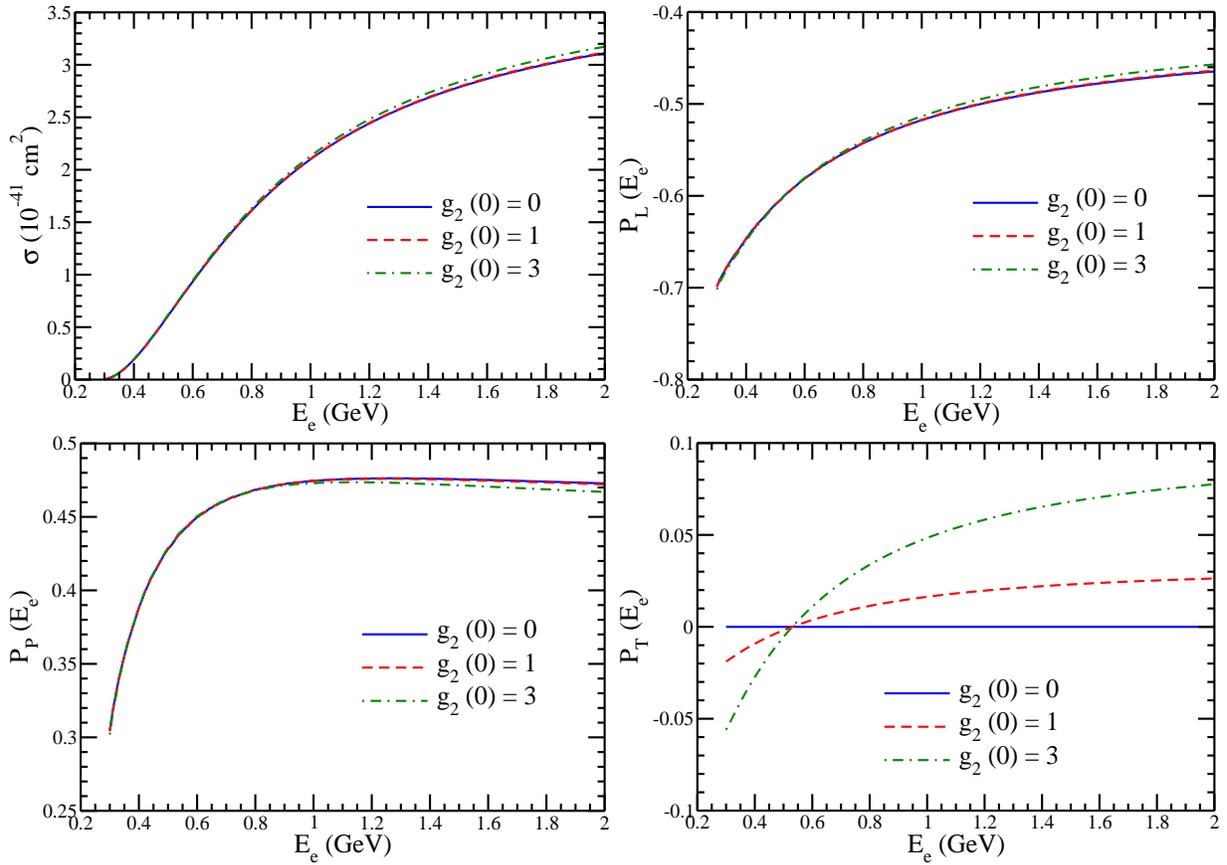

 \includegraphics[height=5.7cm,width=8cm]{total_sigma_g2_variation_Sigma0.eps}
 \includegraphics[height=5.7cm,width=8cm]{Pl_elep_g2_variation_Sigma0.eps}  \\
 \includegraphics[height=5.7cm,width=8cm]{Pp_elep_g2_variation_Sigma0.eps}
 \includegraphics[height=5.7cm,width=8cm]{Pt_elep_g2_variation_Sigma0.eps}
\caption{$\sigma(E_e)~ vs. ~ E_e$~(upper left panel), $P_L (E_e) ~ vs. ~ E_e$~(upper right panel), $P_P (E_e) ~ vs. ~ E_e$~(lower 
left panel) and $P_T (E_e)~ vs.~ E_e$~(lower right panel) for the process ${e^- + p \rightarrow \nu_e + \Sigma^0}$ at $g_2 (0)$ = 
0~(solid line), 1~(dashed line) and 3~(dashed-dotted line) with $M_A = 1.026$ GeV.}\label{fig3}
\end{figure}

\begin{figure}
 \includegraphics[height=5.7cm,width=8cm]{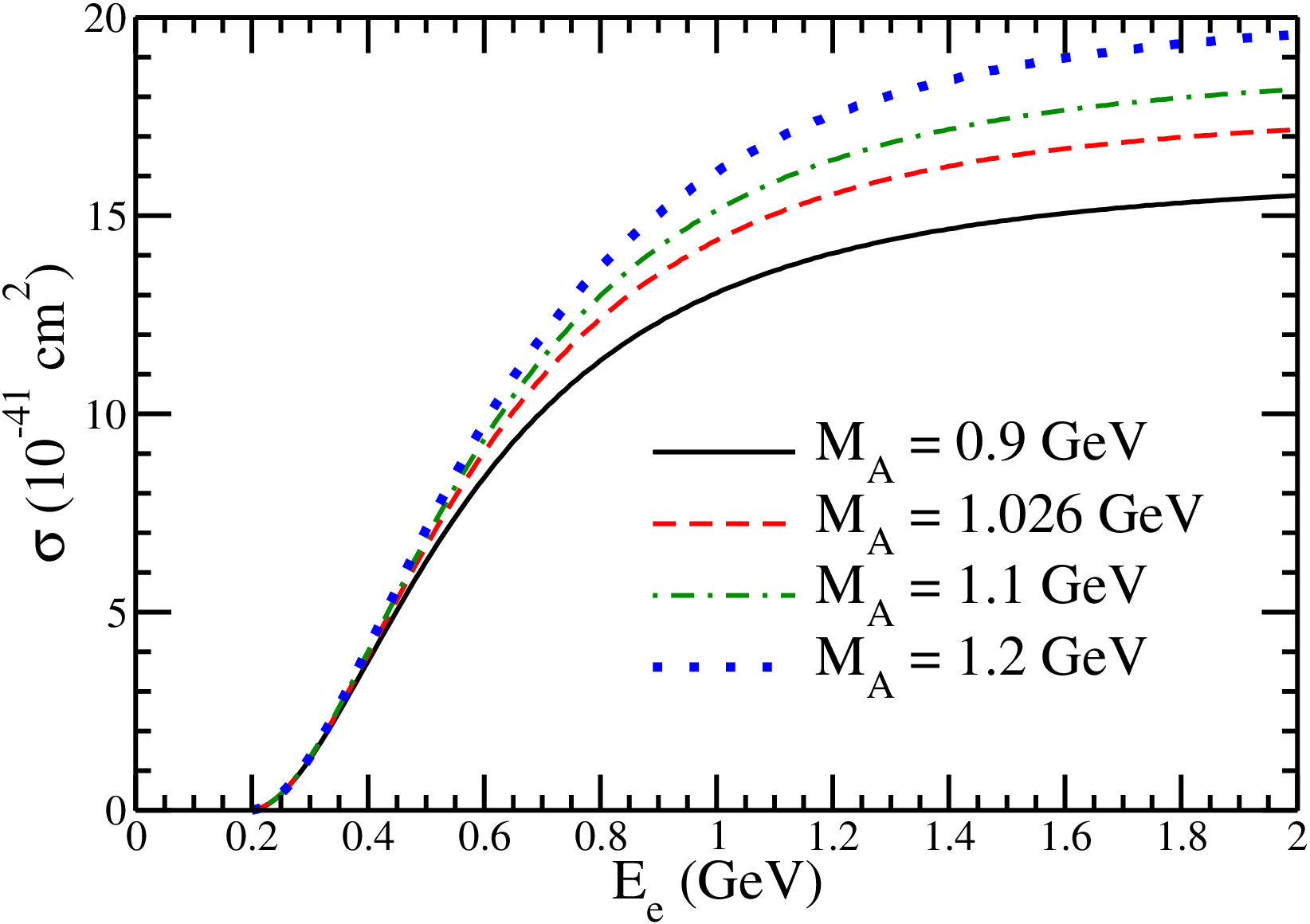}
 \includegraphics[height=5.7cm,width=8cm]{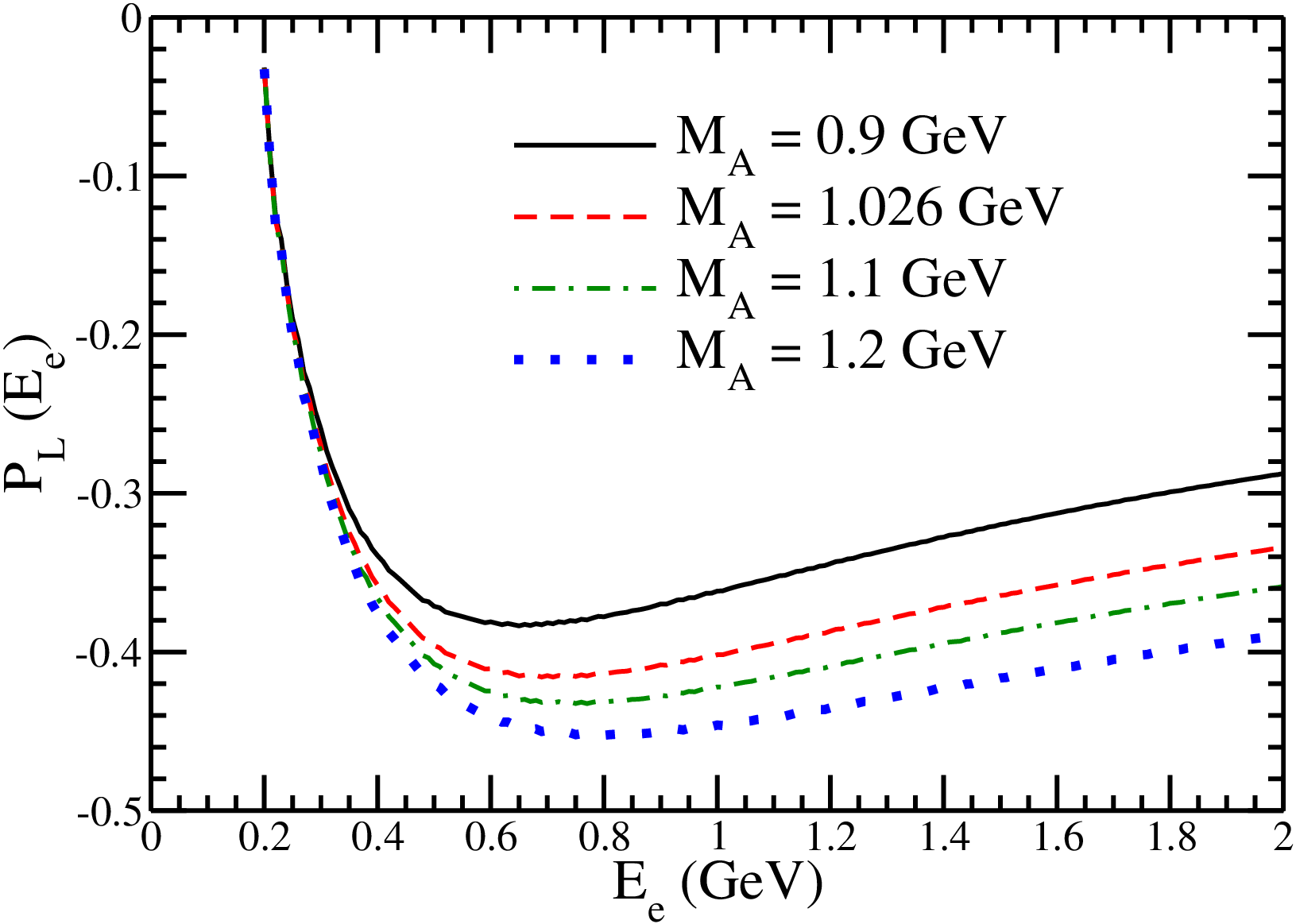}  \\
 \includegraphics[height=5.7cm,width=8cm]{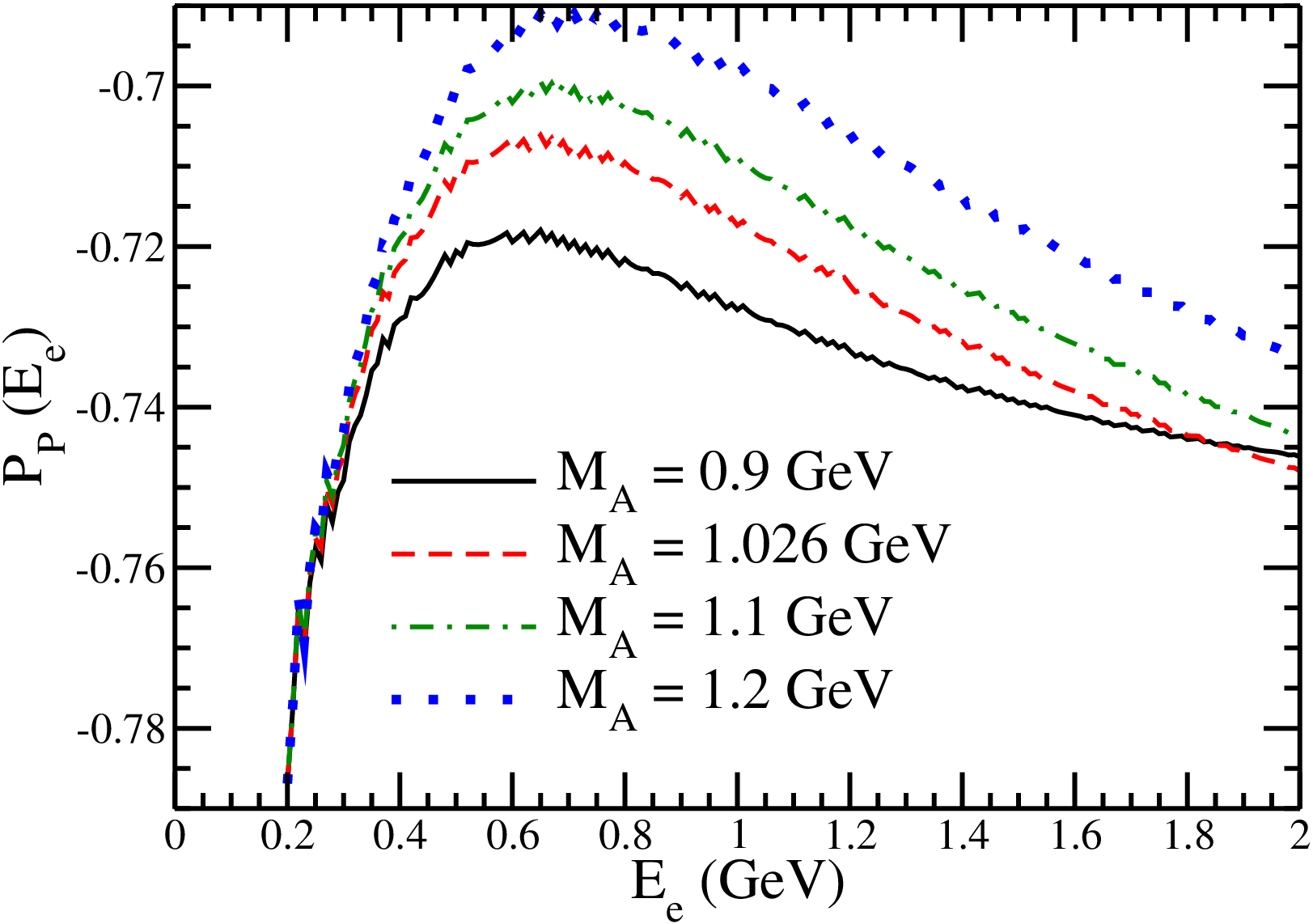}
 \includegraphics[height=5.7cm,width=8cm]{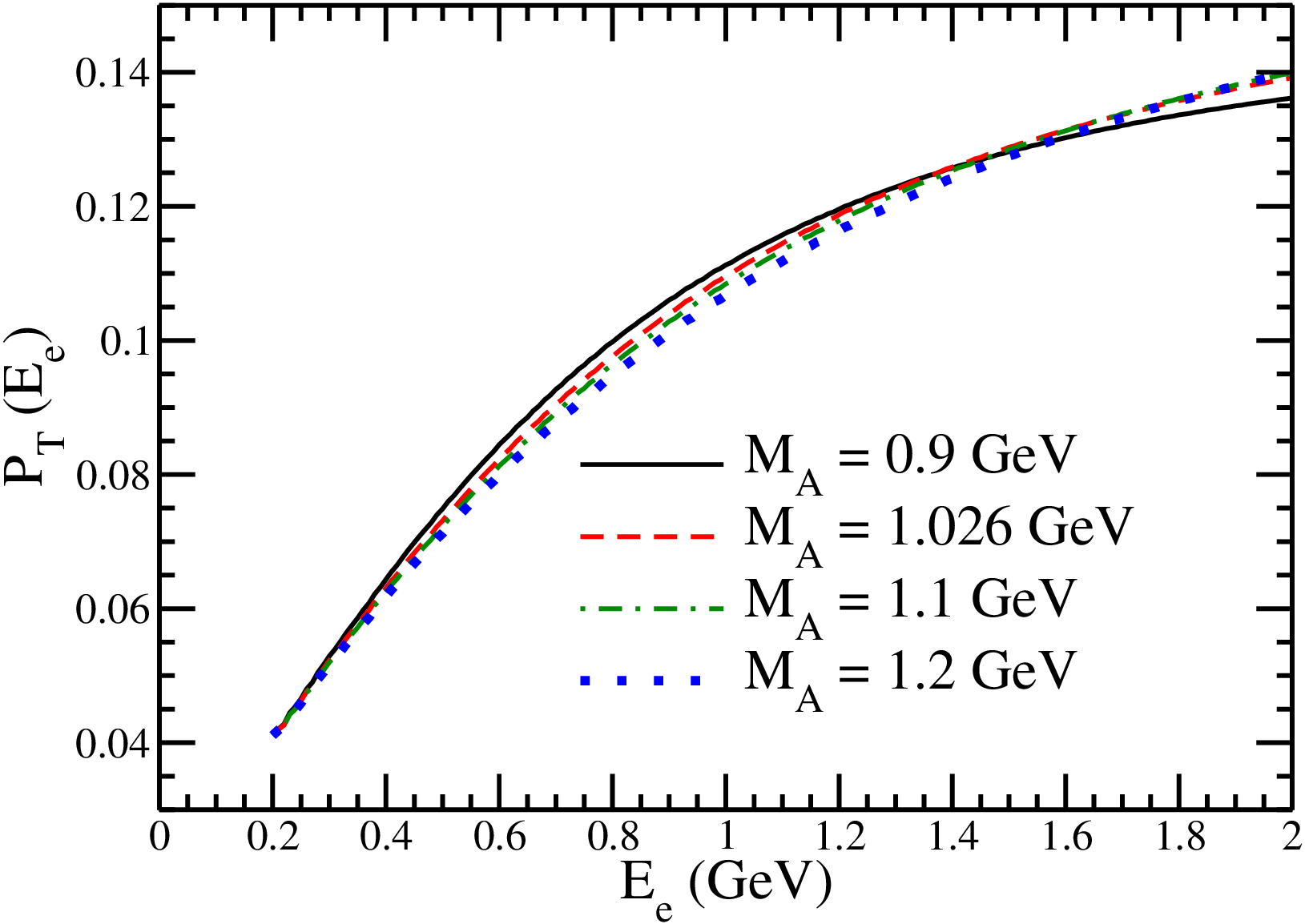}
\caption{$\sigma(E_e)~ vs. ~ E_e$~(upper left panel), $P_L (E_e) ~ vs. ~ E_e$~(upper right panel), $P_P (E_e) ~ vs. ~ E_e$~(lower 
left panel) and $P_T (E_e)~ vs.~ E_e$~(lower right panel) for the process ${e^- + p \rightarrow \nu_e + \Lambda}$ at the different 
values of the axial dipole mass, $M_A = 0.9$ GeV~(solid line), 1.026 GeV~(dashed line), 1.1 GeV~(dashed-dotted line) and 1.2 
GeV~(dotted line) and $g_2 (0) = 1$.}\label{fig5}
\end{figure}
 
\begin{figure}
 \includegraphics[height=6cm,width=8cm]{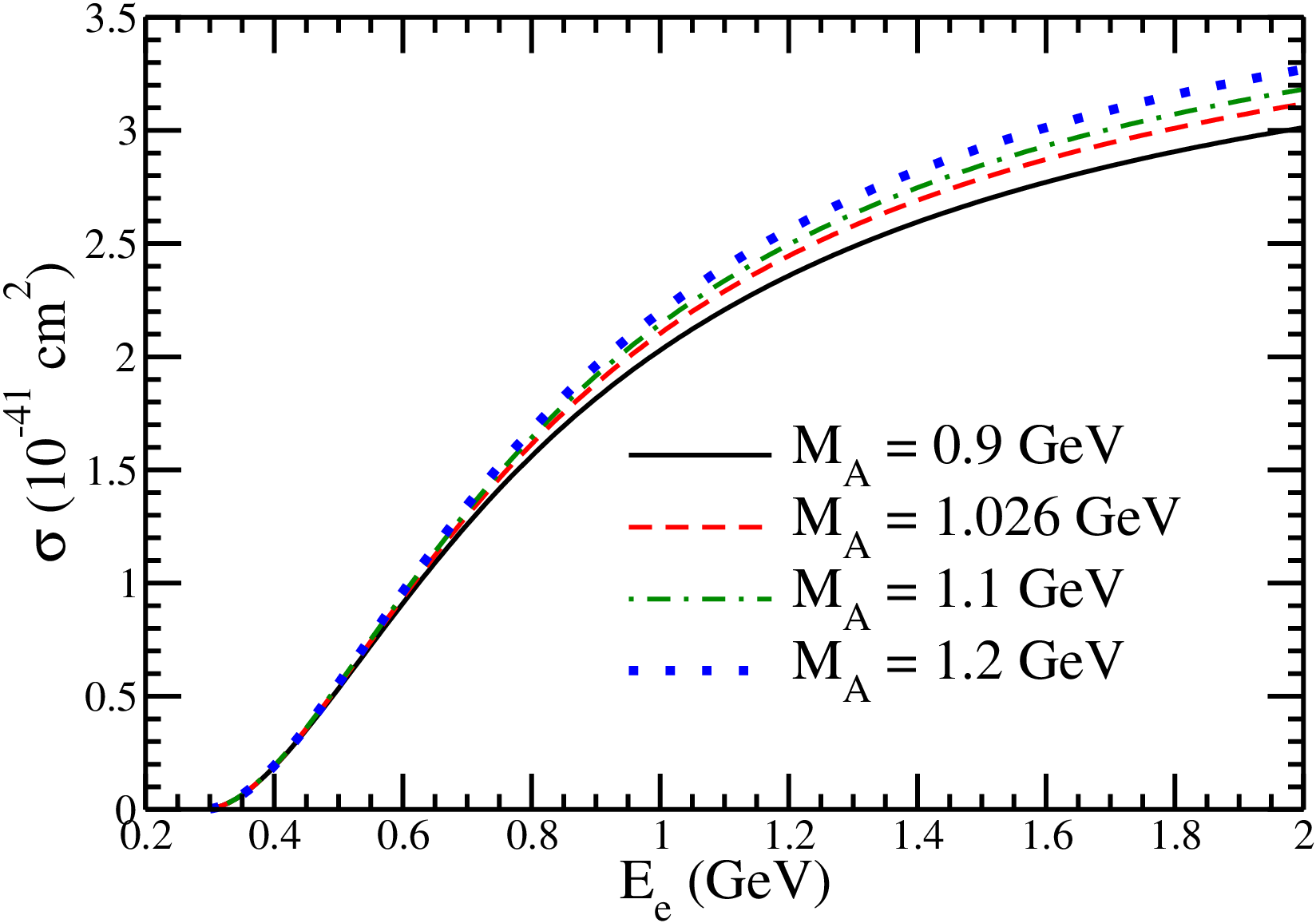}
 \includegraphics[height=6cm,width=8cm]{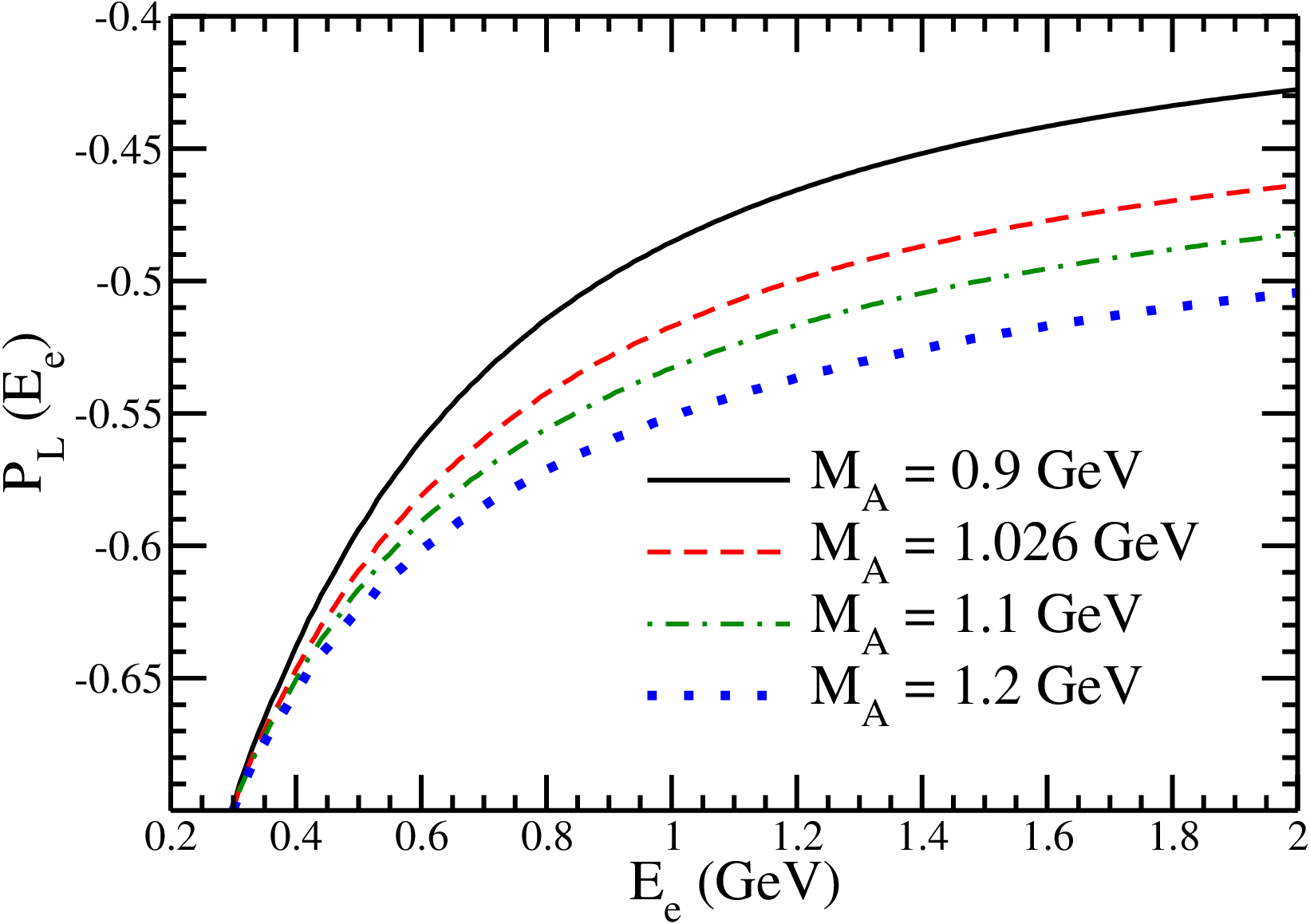}  \\
 \includegraphics[height=6cm,width=8cm]{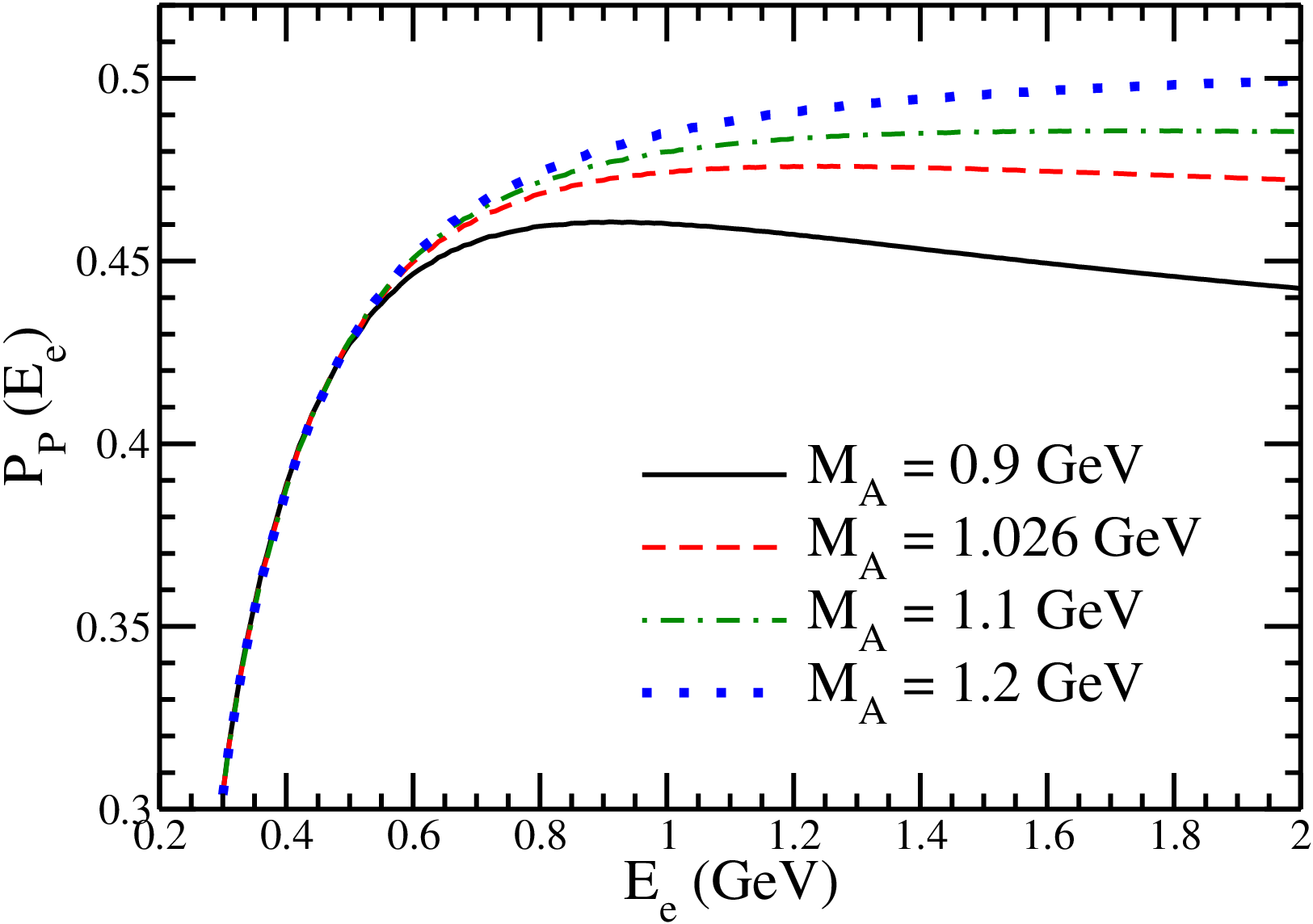}
 \includegraphics[height=6cm,width=8cm]{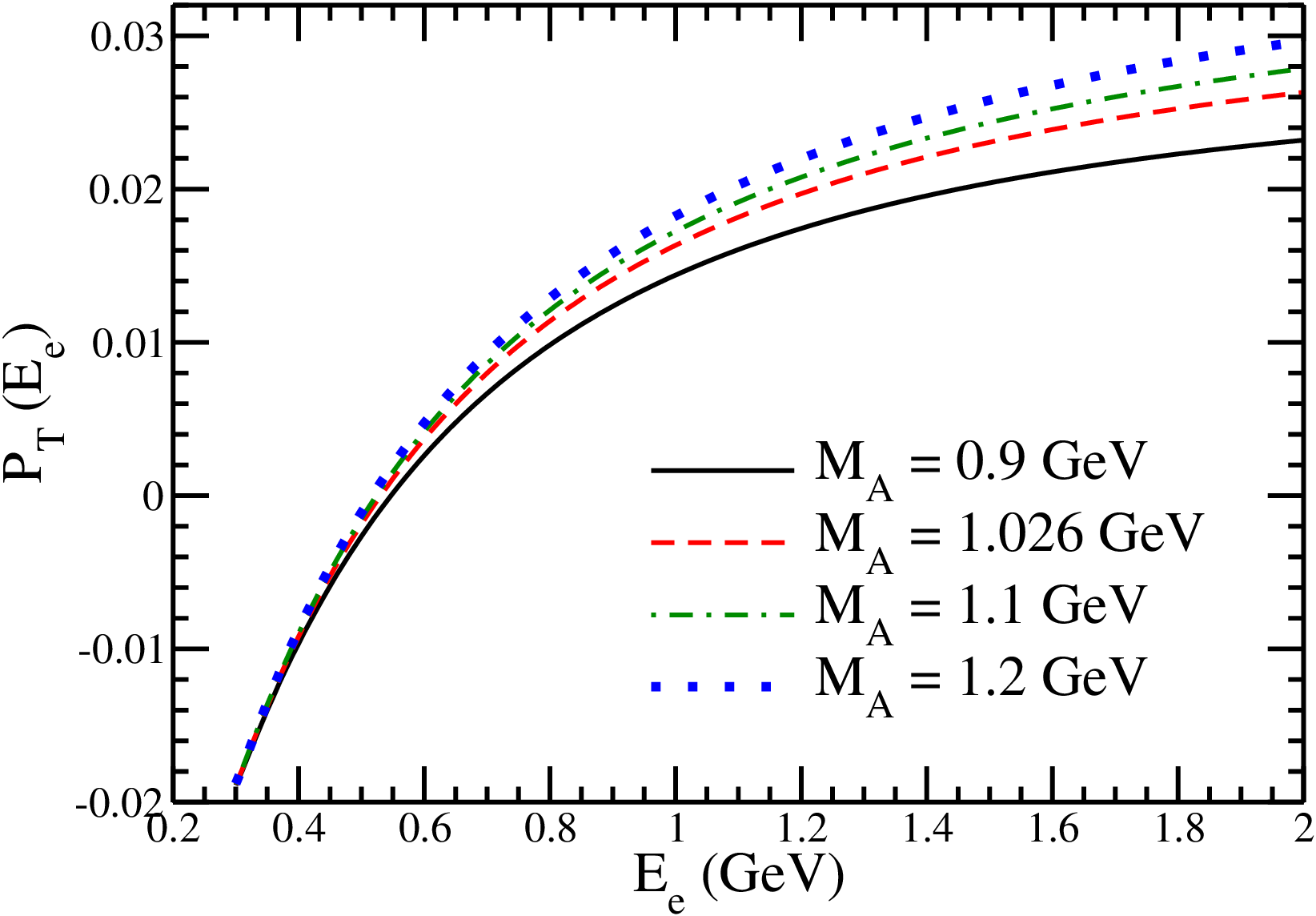}
\caption{$\sigma(E_e)~ vs. ~ E_e$~(upper left panel), $P_L (E_e) ~ vs. ~ E_e$~(upper right panel), $P_P (E_e) ~ vs. ~ E_e$~(lower 
left panel) and $P_T (E_e)~ vs.~ E_e$~(lower right panel) for the process ${e^- + p \rightarrow \nu_e + \Sigma^0}$ at the different 
values of the axial dipole mass, $M_A = 0.9$ GeV~(solid line), 1.026 GeV~(dashed line), 1.1 GeV~(dashed-dotted line) and 1.2 
GeV~(dotted line) and $g_2 (0) = 1$.}\label{fig6}
\end{figure}

\section{Results and discussion}\label{results}
We have used Eq.~(\ref{dsig}) for calculating $d\sigma/dQ^2$ and Eqs.~(\ref{Pl}), (\ref{Pp}) and (\ref{Pt}) for calculating 
$P_L(Q^2)$, $P_P(Q^2)$ and $P_T (Q^2)$. The vector nucleon form factors $f_1^{p,n}(Q^2)$ and $f_2^{p,n} (Q^2)$ which are expressed in 
terms of the electric and the magnetic Sachs form factors, the form for which have been taken from Bradford 
\textit{et al.}~\cite{Bradford:2006yz}. For the axial vector nucleon form factor $g_1 (Q^2)$ and the weak electric form 
factor $g_2^{pn} (Q^2)$, the dipole parameterizations given in Eqs.~(\ref{g1}) and (\ref{g2}) have been used with 
\begin{equation}\label{g22}
 g_2^{pn} (0) = i g_2(0), ~~~ M_2 = M_A.
\end{equation}
To see the dependence of the total cross section $\sigma(E_e)$ and the average polarization observables on the electron's energy, we have 
integrated $d\sigma/dQ^2$, $P_L (Q^2), ~P_P (Q^2)$ and $P_T (Q^2)$ over $Q^2$ using the following expressions: 
\begin{eqnarray}
\sigma(E_e) &=& \int_{Q^2_{min}}^{Q^2_{max}} \frac{d\sigma}{dQ^2} dQ^2, 
\end{eqnarray}
and 
\begin{eqnarray}
 P_{L,P,T} (E_e) &=& \frac{\int_{Q^2_{min}}^{Q^2_{max}} P_{L,P,T} (Q^2) \frac{d\sigma}{dQ^2} dQ^2}{\int_{Q^2_{min}}^{Q^2_{max}} 
 \frac{d\sigma}{dQ^2} dQ^2}
\end{eqnarray}
and show our numerical results for these observables in Figs.~\ref{fig1}--\ref{fig6}. We would like to note that a negative value of $g_2 (0)$ 
(in Eq.~\ref{g22}) does not affect the T-invariant quantities like $\sigma(E_e), ~ P_L(E_e)~ \text{and}~ P_P(E_e)$, but changes the sign 
without changing the magnitude for T-violating variable $P_T(E_e)$.

\begin{figure}
 \includegraphics[height=6cm,width=8cm]{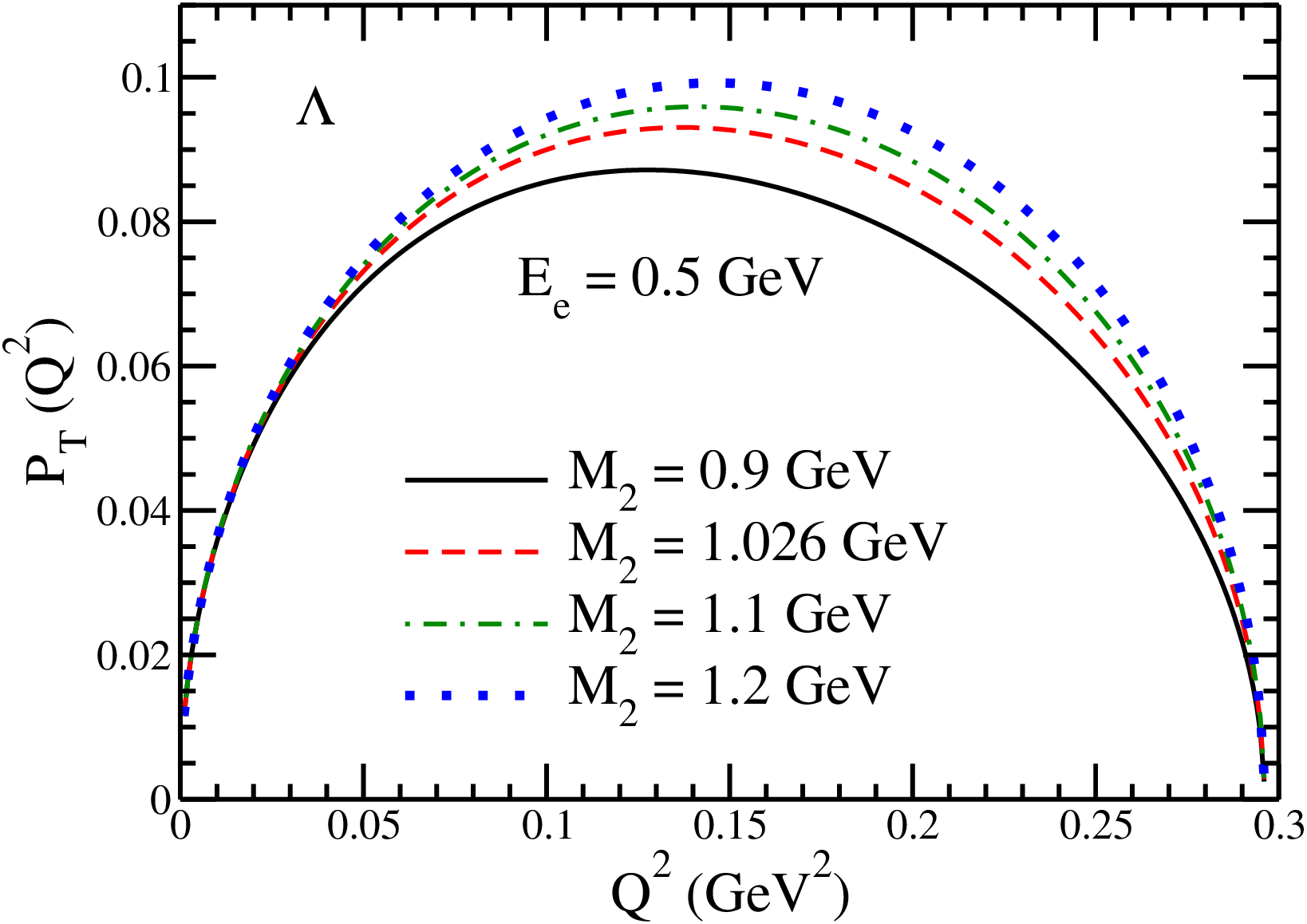}
 \includegraphics[height=6cm,width=8cm]{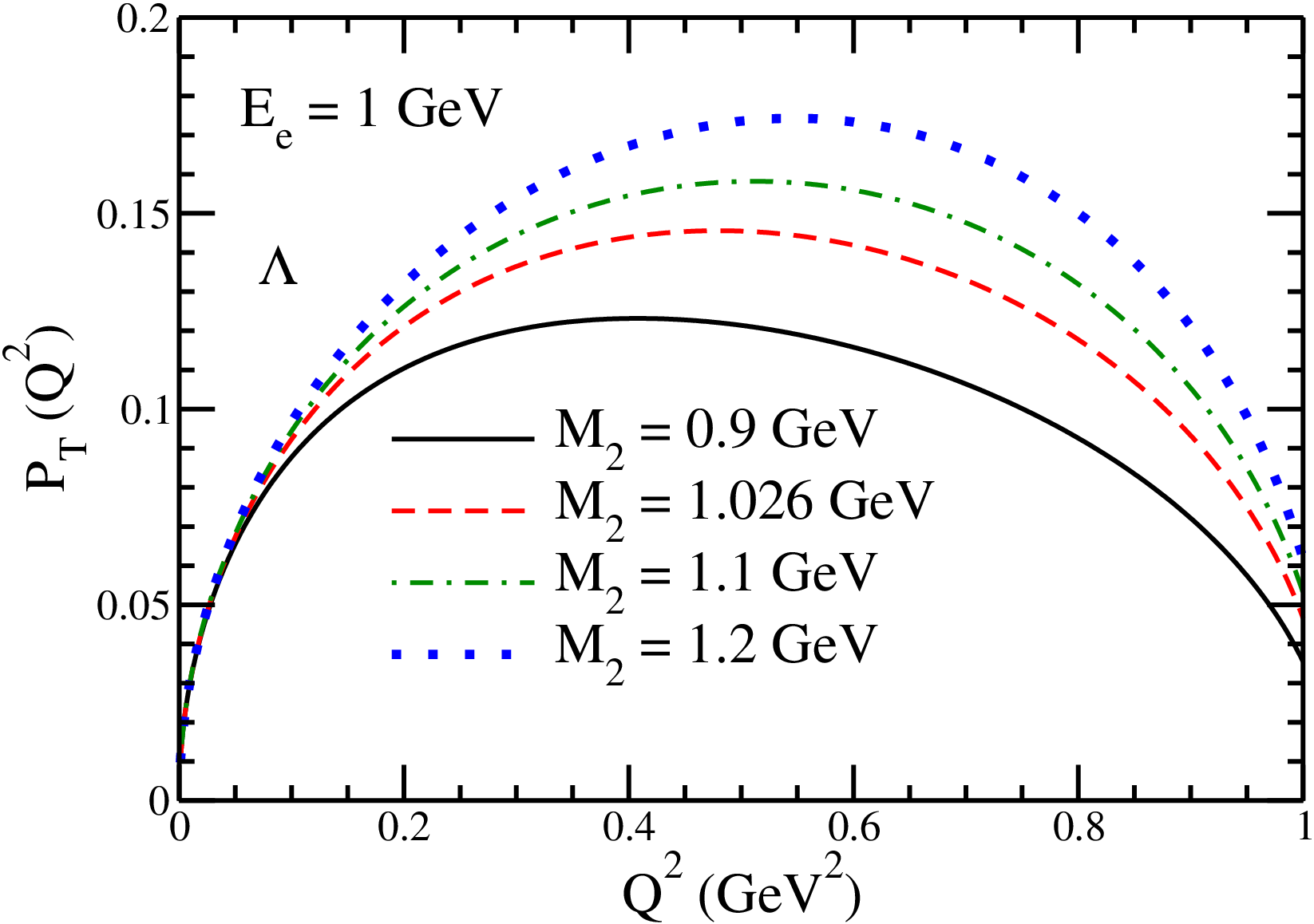}
\caption{$P_T (Q^2)~ vs.~ Q^2$ for the process ${e^- + p \rightarrow \nu_e + \Lambda}$ at the different $E_e$ viz. $E_e =$ 0.5 
GeV~(left panel) and 1 GeV~(right panel) with $M_A = 1.026$ GeV, $g_2 (0)$ = 1 and $M_2 = $ 0.9 GeV~(solid line), 
1.026 GeV~(dashed line), 1.1 GeV~(dashed-dotted line) and 1.2 GeV~(dotted line).}\label{fig7}
\end{figure}

\begin{figure}
 \includegraphics[height=6cm,width=8cm]{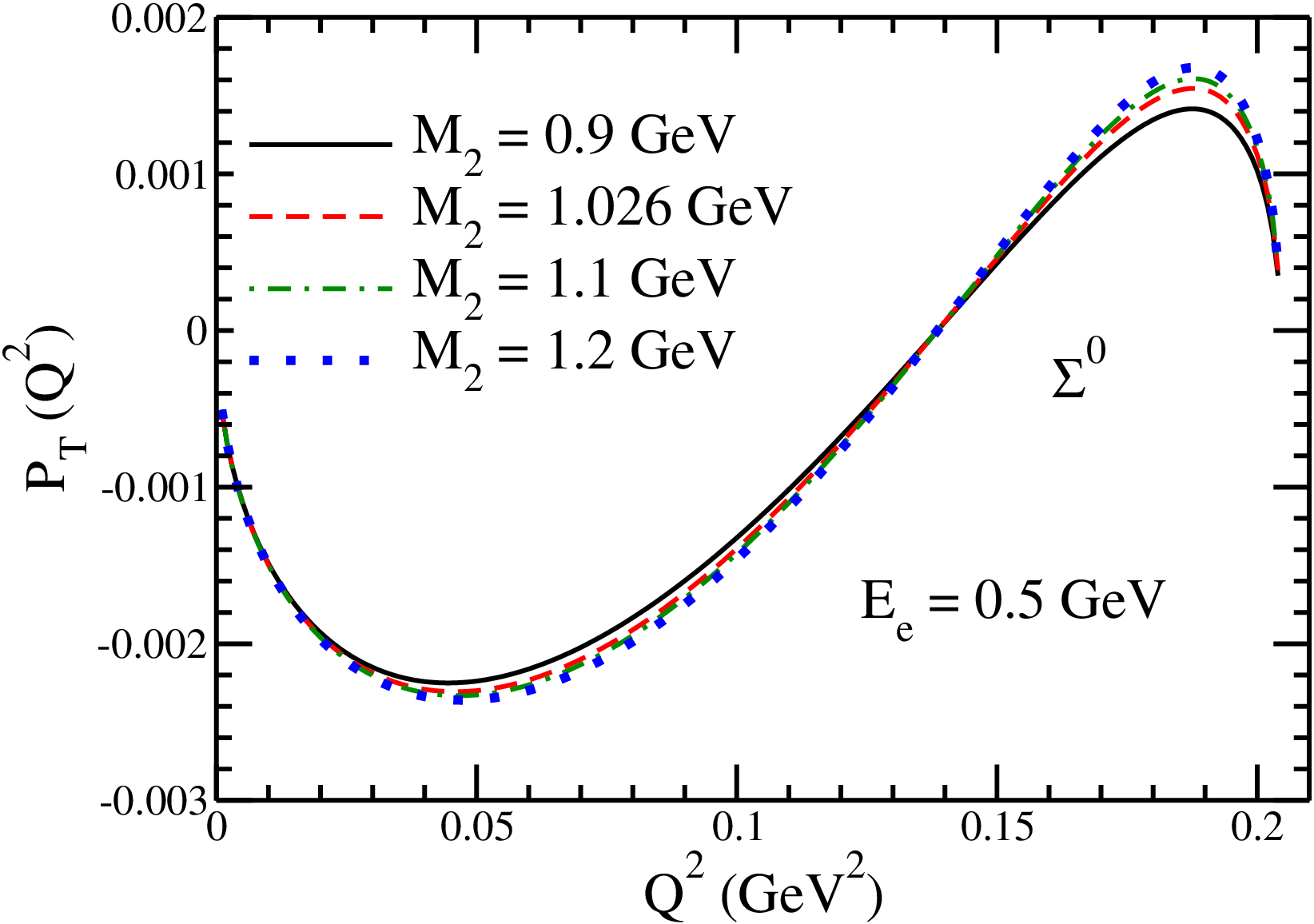}
 \includegraphics[height=6cm,width=8cm]{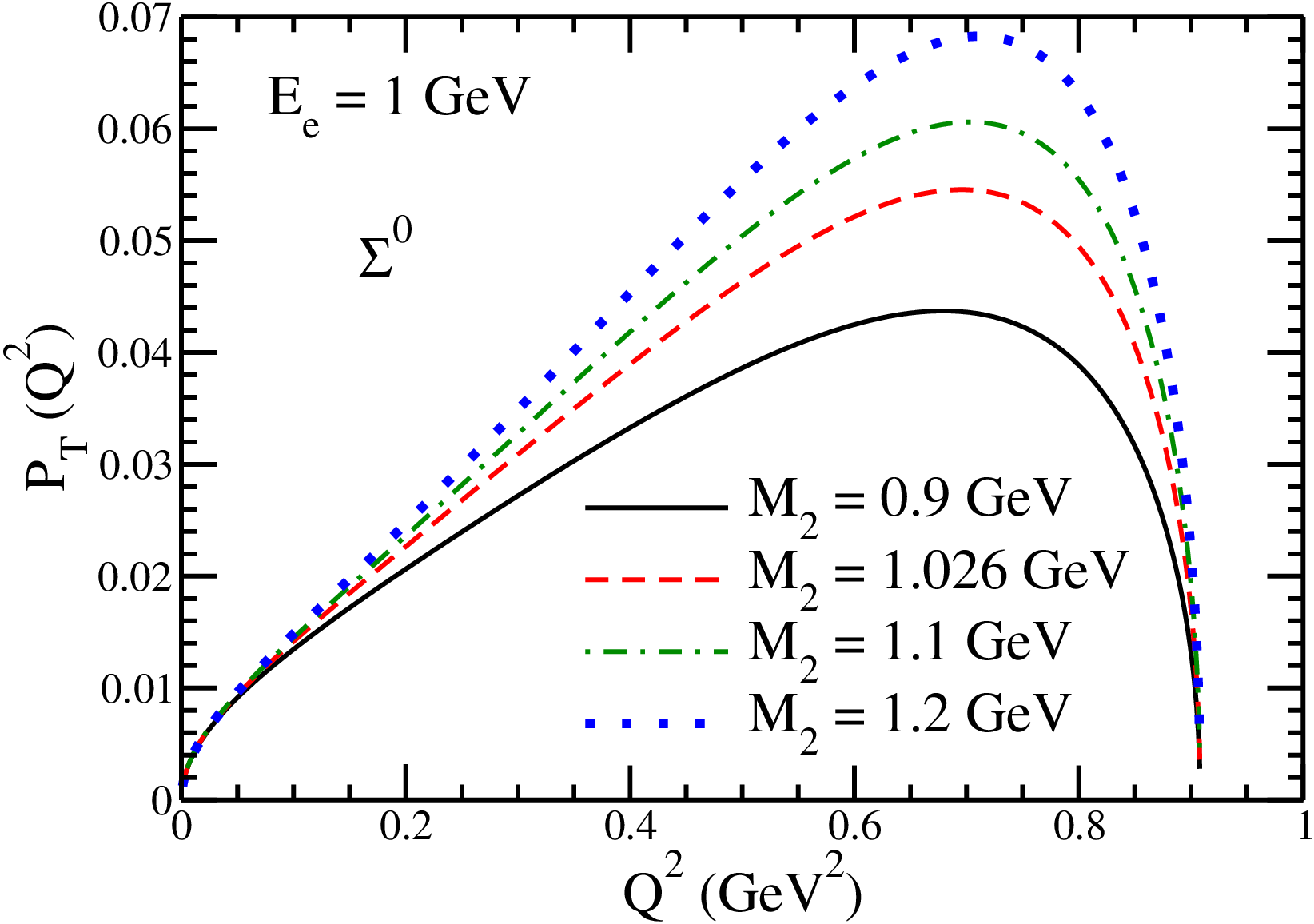}
\caption{$P_T (Q^2)~ vs.~ Q^2$ for the process ${e^- + p \rightarrow \nu_e + \Sigma^0}$ at the different $E_e$ viz. $E_e =$ 0.5 
GeV~(left panel) and 1 GeV~(right panel) with $M_A = 1.026$ GeV, $g_2 (0)$ = 1 and $M_2 = $ 0.9 GeV~(solid line), 
1.026 GeV~(dashed line), 1.1 GeV~(dashed-dotted line) and 1.2 GeV~(dotted line).}\label{fig8}
\end{figure}

\begin{figure}
 \includegraphics[height=6cm,width=8cm]{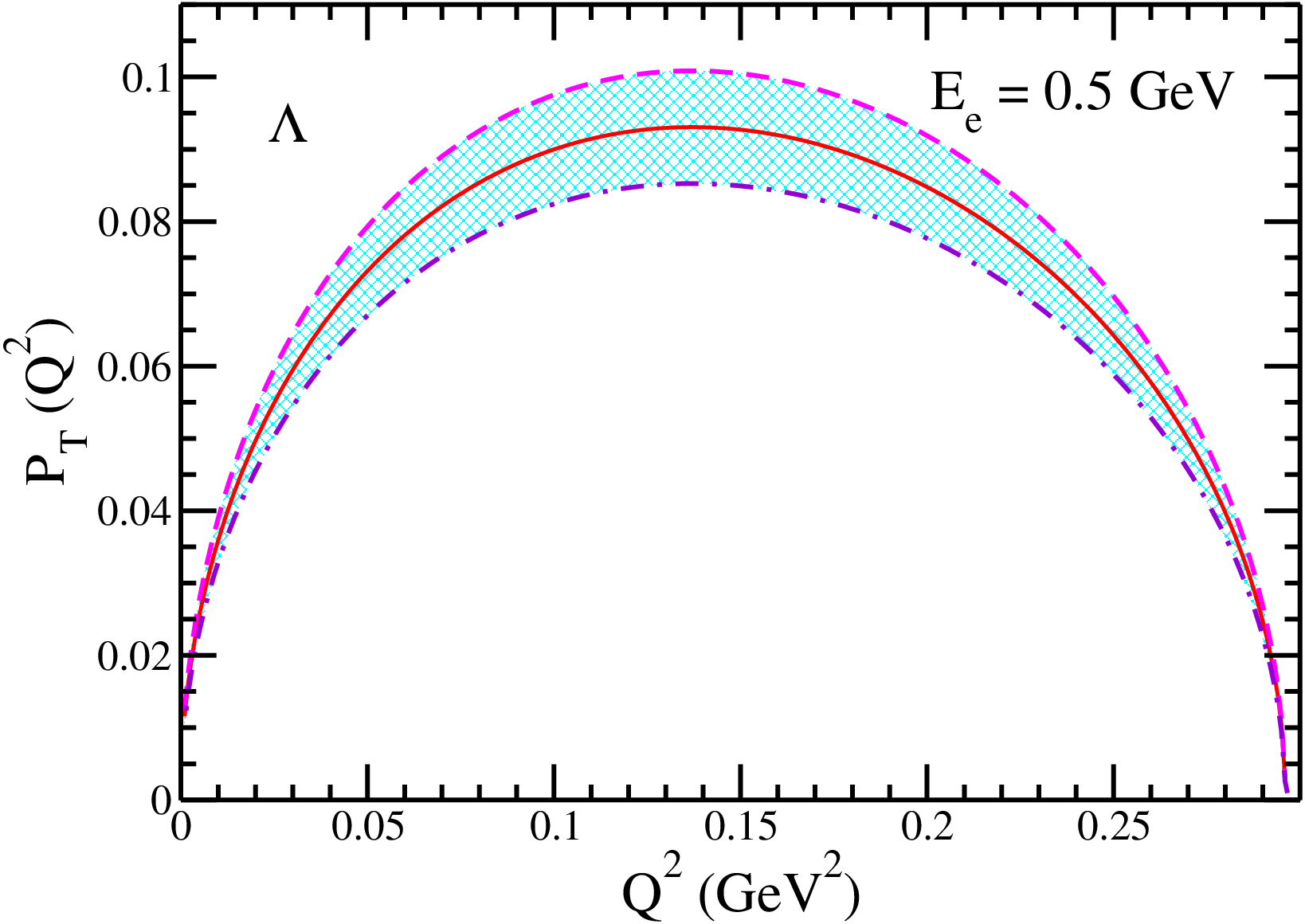}
 \includegraphics[height=6cm,width=8cm]{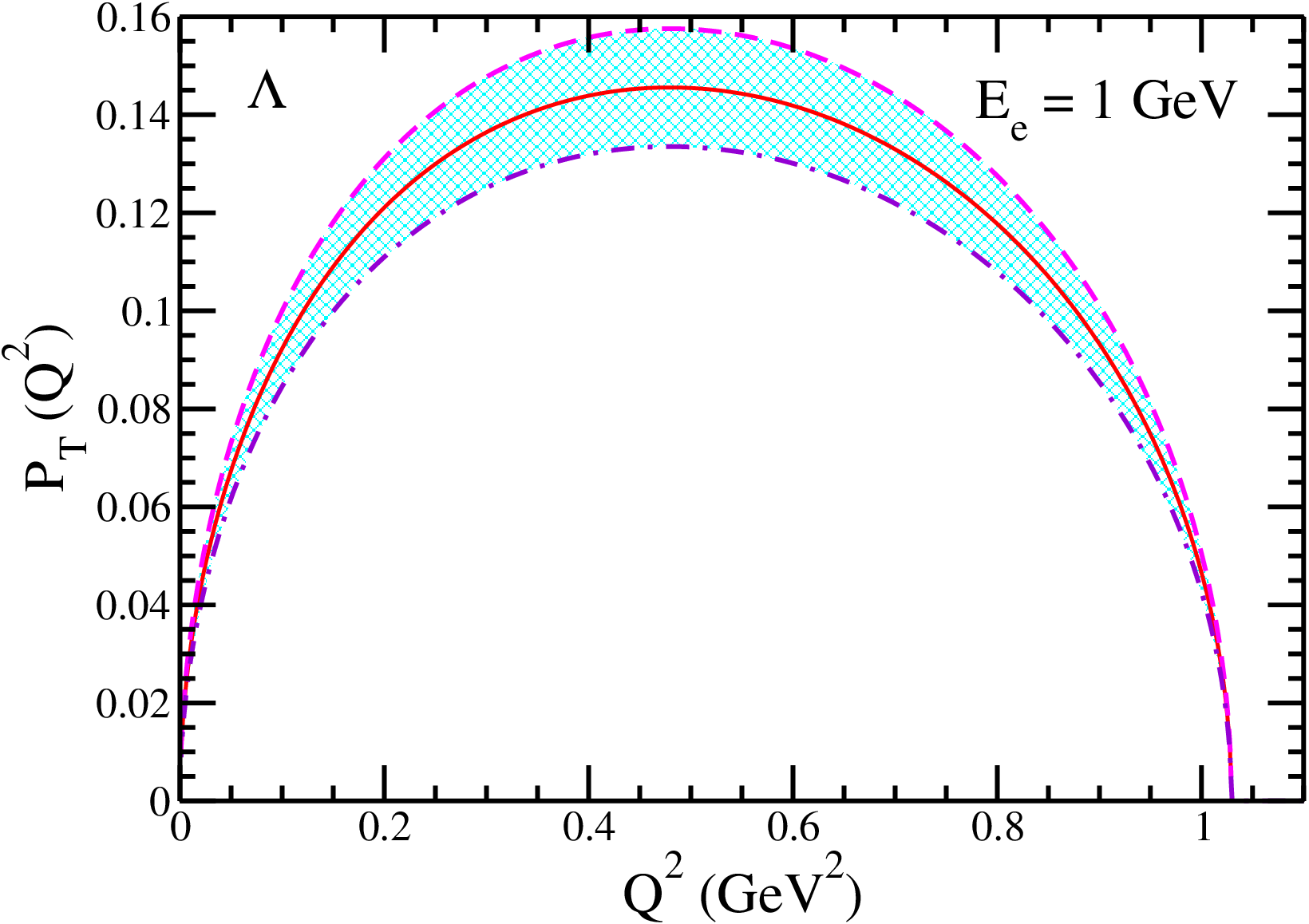}
\caption{$P_T (Q^2)~ vs.~ Q^2$ for the process ${e^- + p \rightarrow \nu_e + \Lambda}$ at the different $E_e$ viz. $E_e =$ 0.5 
GeV~(left panel) and 1 GeV~(right panel) with $M_A = 1.026$ GeV, $g_2 (0)$ = 1, $M_2 = M_A$ and with $x_2 = x_1 (= 0.364)$(solid line),  
$x_2 = x_1 + 20\%~\text{of}~x_1$(dashed line) and $x_2 = x_1 - 20\%~\text{of}~x_1$(dashed-dotted line). The shaded area shows the 
variation of $P_T(Q^2)$ by changing $x_2$ by $\pm 20\%$ around $x_1$.}\label{fig9}
\end{figure}

\begin{figure}
 \includegraphics[height=6cm,width=8cm]{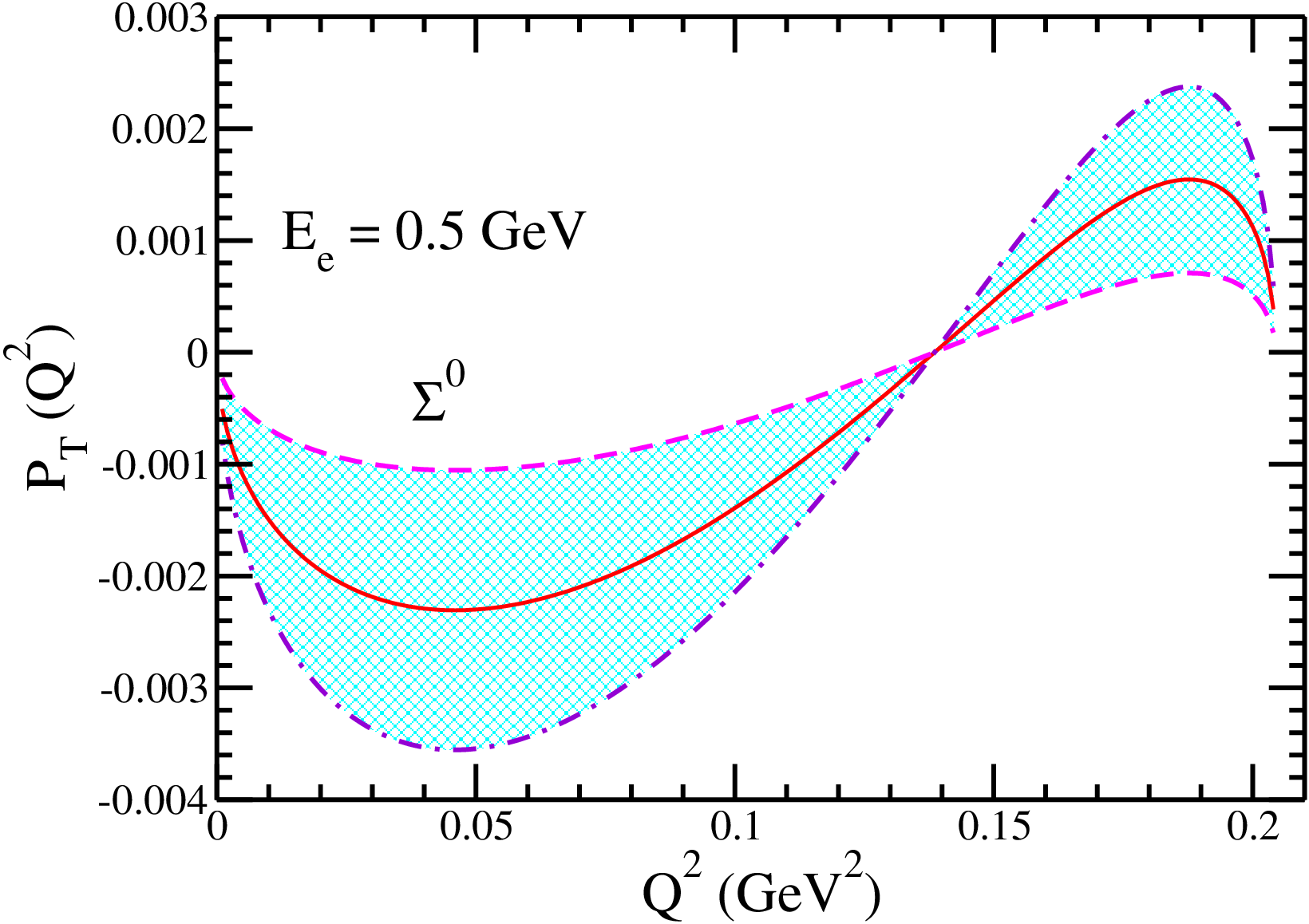}
 \includegraphics[height=6cm,width=8cm]{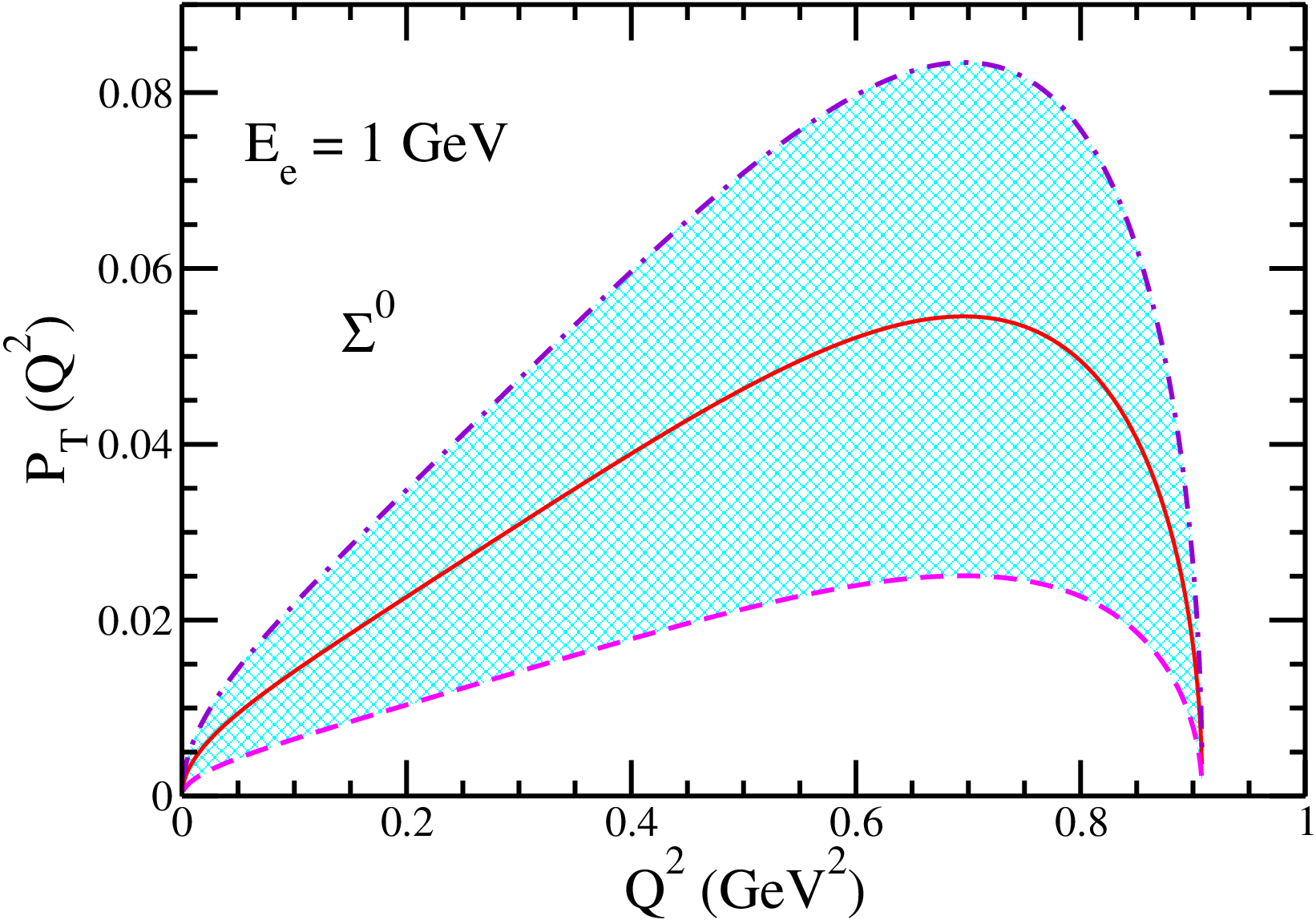}
\caption{$P_T (Q^2)~ vs.~ Q^2$ for the process ${e^- + p \rightarrow \nu_e + \Sigma^0}$ at the different $E_e$ viz. $E_e =$ 0.5 
GeV~(left panel) and 1 GeV~(right panel) with $M_A = 1.026$ GeV, $g_2 (0)$ = 1, $M_2 = M_A$ and with $x_2 = x_1(=0.364)$(solid line),  
$x_2 = x_1 + 20\%~\text{of}~x_1$(dashed line) and $x_2 = x_1 - 20\%~\text{of}~x_1$(dashed-dotted line). The shaded area shows the 
variation of $P_T(Q^2)$ by changing $x_2$ by $\pm 20\%$ around $x_1$.}\label{fig10}
\end{figure}

In Fig.~\ref{fig1}, we have presented the results for $\sigma(E_e)$ and $P_L (E_e)$, $P_P (E_e)$ and $P_T (E_e)$ for neutron, as a 
function of $E_e$ for the process $e^- + p \rightarrow \nu_e + n$ by taking $g_2 (0)$ = 0, 1 and 3, and $M_2 (= M_A) = $1.026 GeV 
in Eqs.~(\ref{g1}) and (\ref{g2}). We find that there is no appreciable $g_2 (Q^2)$ dependence on the total scattering cross section 
as well as on the polarization observables $P_L (E_e)$ and $P_P (E_e)$ unless $g_2(0) \ge 2$. However, the transverse polarization 
$P_T(E_e)$ could be between $10-30\%$ at $E_e = 1$ GeV for $1\le g_2(0) \le 3$ and larger for $E_e \ge 1$ GeV. In view of the 
experimental difficulties in measuring the neutron polarization~\cite{Block:1965zol, Block:NAL, Javannic:NAL}, we do not discuss it 
further and limit our results and discussions for the case of the hyperon production where the polarization measurements are 
relatively easier as the pionic decays of hyperons are self analyzing. 

In Fig.~\ref{fig2}, we have presented the results for $\sigma(E_e)$, $P_L (E_e)$, $P_P (E_e)$ and $P_T (E_e)$ for the process 
$e^- + p \rightarrow \nu_e + \Lambda$ using $g_2 (0)$ = 0, 1 and 3 and $M_2 (= M_A) = $1.026 GeV. We find that there is very 
little dependence of $\sigma(E_e)$ and $P_L (E_e)$ on $g_2(0)$, $P_P (E_e)$ and $P_T(E_e)$ have appreciable dependence on $g_2(0)$ 
specially if $g_2(0) \ge 1$. In Fig.~\ref{fig3}, we have presented the results for $\sigma(E_e)$, $P_L (E_e)$, $P_P (E_e)$ and 
$P_T (E_e)$ for the process $e^- + p \rightarrow \nu_e + \Sigma^0$ at $g_2 (0)$ = 0, 1 and 3. For $\sigma(E_e)$, $P_L (E_e)$ and 
$P_P (E_e)$, we find almost no dependence on $g_2 (0)$ while for $P_T(E_e)$ it could be $2-8\%$ for $1\le g_2(0)\le 3$.

Thus, it is possible to study the effect of T-noninvariance in the hyperon production reactions induced by the electrons, i.e., 
$e^- + p \longrightarrow \nu_e + \Lambda(\Sigma^0)$, provided the T-violating form factor $|g_2(0)| \ge 1$.
 
To observe the dependence of $\sigma(E_e)$, $P_L (E_e)$, $P_P (E_e)$ and $P_T (E_e)$ on $M_A$, we have varied $M_A$ in the range 
$0.9-1.2$ GeV, and obtained the results which are shown in Fig.~\ref{fig5} for the process $e^- + p \rightarrow \nu_e + \Lambda$ with 
$g_2 (0) = 1$ and $M_2 = $1.026 GeV. It may be observed from the figure that while $\sigma(E_e)$, $P_L (Q^2)$ and $P_P(Q^2)$ are 
sensitive to the value of $M_A$, $P_T(E_e)$ shows almost no $M_A$ dependence. In the case a higher value of $g_2(0)$ is chosen, 
the $M_A$ dependence of $\sigma(E_e),~ P_{L,P,T} (E_e)$ is qualitatively the same except that the absolute values of these 
observables, i.e., $\sigma(E_e),~ P_{L,P,T} (E_e)$ are quantitatively larger. For example, with $g_2(0) = 3$ at $E_e = 1$ GeV, 
$\sigma(E_e),~ P_L(E_e), ~ P_P(E_e)$ are enhanced by $\sim$ 6$\%$, 4$\%$ and 8$\%$, respectively, while $P_T(E_e)$ is enhanced 
almost by a factor of 1.8.

  In Fig.~\ref{fig6}, $\sigma(E_e)$, $P_L (E_e)$, $P_P (E_e)$ and $P_T(E_e)$ for the process 
 $e^- + p \rightarrow \nu_e + \Sigma^0$ at the different values of the axial dipole mass, $M_A$ = 0.9, 1.026, 1.1 and 1.2 GeV for 
 $g_2 (0) = 1$ are presented. The results in this case are qualitatively similar to those presented in Fig.~\ref{fig5} for 
 $e^- + p \rightarrow \nu_e + \Lambda$. 
 
 We have also studied the dependence of $d\sigma/dQ^2$, $P_L (Q^2)$, $P_P (Q^2)$ and $P_T (Q^2)$ on $M_2$ at the two different 
 energies of the electron, viz. $E_e =$ 0.5 GeV and 1 GeV by taking the values $M_2 = $ 0.9, 1.026 GeV, 1.1 GeV and 1.2 GeV and 
 $g_2(0) = 1$. We find that the results for $d\sigma/dQ^2$, $P_L (Q^2)$ and $P_P (Q^2)$ are insensitive to the variation in the 
 value of $M_2$ (not shown) whereas the results for $P_T(Q^2)$ show some dependence of $M_2$. The results are shown in 
 Fig.~\ref{fig7} for $e^- + p \rightarrow \nu_e + \Lambda$ and in Fig~\ref{fig8} for ${e^- + p \rightarrow \nu_e + \Sigma^0}$. 
 
 We have also studied the effect of varying $x_2$ by $20\%$ around the value of $x_1 = 0.364$ on the differential cross section ($d \sigma/d Q^2$)
 and the polarization observables $P_L(Q^2)$, $P_P(Q^2)$ and $P_T(Q^2)$, and found that $d \sigma/d Q^2$, $P_L(Q^2)$ and $P_P(Q^2)$ are insensitive to 
 the variation in $x_2$ (not shown here). We present our results for $P_T(Q^2)$ in Figs.~\ref{fig9} and \ref{fig10} for 
 $e^- + p \longrightarrow \Lambda + \nu_e$ and $e^- + p \longrightarrow \Sigma^0 + \nu_e$ processes respectively.
 
  \section{Conclusion}\label{conclusion}
  We have studied in this work the differential scattering cross section, total scattering cross section as well as the longitudinal, 
  perpendicular and transverse components of the polarization for n, $\Lambda$ and $\Sigma^0$ produced in the quasielastic reaction 
  of the electron on the free proton target induced by the weak charged current. The form factors for the nucleon-hyperon transition 
  have been obtained using the Cabibbo theory assuming SU(3) symmetry, CVC, PCAC. For the weak electric form factor 
  $g_2^{pn}(Q^2)$, we have taken dipole form with dipole mass $M_2 = M_A$ = 1.026 GeV and $g_2^{pn}(0)$ to be purely imaginary, i.e., 
  $g_2^{pn}(0) = i g_2(0)$ and have varied $g_2(0)$ in the range $0 \le |g_2(0)| \le 3$.  We have studied the dependence of the cross 
  section and the polarization observables on the weak transition form factor $g_1^{pX} (Q^2)$ and $g_2^{pX} (Q^2)$ with $X=n, 
  \Lambda, \Sigma^0$. 
 
 To summarize our results we find that:
  \begin{enumerate}
  \item[1)] The consideration of the T-noninvariance terms in the matrix element of the weak hadronic currents leads to the 
  non-vanishing transverse polarization $P_T(Q^2)$ of the final hadron in a direction perpendicular to the plane of reaction. The 
  transverse polarization $P_T(Q^2)$ for $\Lambda$ is large in the case of $e^- + p \longrightarrow \nu_e + \Lambda$ as compared to 
  the $\Sigma^0$ polarization in $e^- + p \longrightarrow \nu_e + \Sigma^0$. Therefore, it is possible to study the physics of 
  T-noninvariance by experimentally measuring this polarization in future experiments.
  
  \item[2)] The longitudinal and perpendicular components of the polarization $P_L(Q^2)$ and $P_P(Q^2)$ are sensitive to the value 
  of axial dipole mass $M_A$, especially in the case of $P_L(Q^2)$ for $\Lambda$ as well as $\Sigma^0$ production. A study of the 
  $Q^2$ dependence of $P_L(Q^2)$ and $P_P(Q^2)$ will enable us to make the measurements for the axial vector form factor independent 
  of the cross section measurements.
  
  \item[3)] The dependence of the differential $d\sigma/dQ^2$ and the total $\sigma(E_e)$ cross section, and the polarization 
  observables $P_L (E_e,Q^2)$, $P_P (E_e,Q^2)$ and $P_T (E_e,Q^2)$ on the T-violating weak electric form factor $g_2(Q^2)$ has been 
  studied by varying $g_2(0)$ and the dipole mass $M_2$. It is found that 
  
  \begin{enumerate}
   \item The total cross section $\sigma(E_e)$ and the polarization components $P_L(E_e)$ and $P_P(E_e)$ are sensitive to $g_2(0)$ 
   only when $g_2(0)\ge 2$ but $P_T(E_e)$ being directly dependent on $g_2(0)$ is quite sensitive to it being non-zero.
   
   \item While the total cross section $\sigma(E_e)$ and the polarization components $P_{L,P}(E_e)$ are sensitive to the axial dipole 
   mass $M_A$, the transverse component of polarization $P_T(E_e)$ is quite insensitive to the variation in $M_A$.
   
   \item The differential cross section $d\sigma/dQ^2$ and the polarization components $P_{L,P}(Q^2)$ have very weak dependence on 
   $M_2$ but the transverse component of the polarization $P_T(Q^2)$ depends significantly on $M_2$.
  \end{enumerate}
  \end{enumerate}

  \section{Acknowledgment}   
M. S. A. and S. K. S. are thankful to Department of Science and Technology (DST), Government of India for providing financial 
assistance under Grant No. EMR/2016/002285.

\section{Appendix}
The expressions $A(E_e,Q^2)$, $B(E_e, Q^2)$, $C(E_e,Q^2)$ and $N(E_e,Q^2)$ are expressed in terms of the Mandlstam variables and the 
form factors as:
\begin{eqnarray}
 A(E_e,Q^2) &=& 2 \Bigg[f_1^2(Q^2) \left(\frac{1}{2} (M + M^\prime) \left((M - M^\prime)^2-t\right) \right) + \frac{f_2^2(Q^2)}
{(M+M^\prime)^2} \left(\frac{1}{2} t (M+M^\prime) \left((M-M^\prime)^2-t\right) \right) \nonumber \\ 
 &+& g_1^2(Q^2) \left(\frac{1}{2} (M-M^\prime) \left(t-(M+M^\prime)^2\right) \right) + \frac{|g_2 (Q^2)|^2}{(M+M^\prime)^2} 
 \left(-\frac{1}{2} t (M - M^\prime) \left((M + M^\prime)^2-t\right) \right) \nonumber \\
 &+& \frac{f_1 (Q^2) f_2 (Q^2)}{(M+M^\prime)} \left(\frac{1}{2} \left( (M^2-{M^\prime}^2)^2-4 M M^\prime t-t^2\right) \right) 
 + f_1(Q^2) g_1(Q^2) \left(-M^\prime  (M^2+ {M^\prime}^2-2 s-t) \right) \nonumber \\
 &+& \frac{Re[f_1(Q^2) g_2(Q^2)]}{(M+M^\prime)} (\frac{1}{2} (M^2-{M^\prime}^2-t) (M^2+{M^\prime}^2-2 s-t) ) 
 \nonumber \\
 &+& \frac{f_2(Q^2) g_1(Q^2)}{(M+M^\prime)} \left(\frac{1}{2} (M^2-{M^\prime}^2-t) (M^2+{M^\prime}^2-2 s-t) \right) \nonumber \\
 &+& \frac{Re[f_2(Q^2) g_2(Q^2)]}{(M+M^\prime)^2} \left(-M^\prime t (M^2+{M^\prime}^2-2 s-t) \right) \nonumber \\
 &+& \left.  \frac{Re[g_1(Q^2) g_2(Q^2)]}{(M+M^\prime)} \left(\frac{1}{2} \left( (M^2-{M^\prime}^2)^2+4 M M^\prime t-t^2\right)
 \right)  \right], \\
 B(E_e,Q^2) &=& \frac{2}{M^\prime} \Bigg[f_1^2(Q^2) \left( \frac{1}{4} \left(-2 M^\prime (M - M^\prime) \left(M^2-s\right)-t 
 \left((M-M^\prime)^2 -2 s\right)+t^2\right)  \right)  \nonumber \\
 &+& \frac{f_2^2(Q^2)}{(M+M^\prime)^2} \left(-\frac{1}{4} t (M + M^\prime) \left(M^3+M^2 M^\prime - M ({M^\prime}^2+2 s+t)
 +{M^\prime}^3  - M^\prime t\right) \right) \nonumber \\
 &+& g_1^2(Q^2) \left(\frac{1}{4} \left(2 M^\prime (M + M^\prime) \left(M^2-s\right)-t \left((M+M^\prime)^2-2 s\right)+t^2\right) 
 \right) \nonumber \\
 &+& \frac{|g_2(Q^2)|^2}{(M+M^\prime)^2} \left(-\frac{1}{4} t (M - M^\prime) \left(M^3-M^2 M^\prime - M ({M^\prime}^2+2 s+t)
 - {M^\prime}^3 + M^\prime t \right) \right) \nonumber \\
 &+& \frac{f_1(Q^2) f_2(Q^2)}{(M+M^\prime)} \left(-\frac{1}{2} \left(M^4 M^\prime + M^3 t - M^2 M^\prime ({M^\prime}^2+s) 
 - M t ({M^\prime}^2+2 s +t) + M^\prime ({M^\prime}^2-t) (s+t)\right) \right) \nonumber \\
 &+& f_1(Q^2) g_1(Q^2) \left(\frac{1}{2} \left(M^2 (2 s+t)+t ({M^\prime}^2-t)-2 s^2-2 s t\right) \right) \nonumber \\
 &+& \frac{Re[f_1(Q^2) g_2(Q^2)]}{(M+M^\prime)} \left(-\frac{1}{2} \left(- m_e^2 (M + M^\prime) (M^2+{M^\prime}^2-t)+
 M^4 M^\prime+M^3 t +M^2 M^\prime ({M^\prime}^2 -3 s-2 t) \right. \right. \nonumber \\
 &+& M t \left. \left. ({M^\prime}^2-t)-M^\prime (s+t) ({M^\prime}^2-2 s-t)\right) \right) \nonumber \\
 &+& \frac{f_2(Q^2) g_1(Q^2)}{(M+M^\prime)} \left(\frac{1}{2} \left(m_e^2 (-(M - M^\prime)) (M^2+{M^\prime}^2-t)-M^\prime (M^2-s) 
 (M^2+ {M^\prime}^2-2 s) \right. \right. \nonumber \\
 &+& t \left. \left. (M^3+2 M^2 M^\prime +M {M^\prime}^2+{M^\prime}^3-3 M^\prime s)-t^2 (M+M^\prime)\right) \right) \nonumber \\
 &+& \frac{Re[f_2(Q^2) g_2(Q^2)]}{(M+M^\prime)^2} \left(\frac{1}{2} t \left(m_e^2 (M^2+{M^\prime}^2-t)-M^4+M^2 (2 s+t)+{M^\prime}^4
 -{M^\prime}^2 t-2 s (s+t)\right) \right) \nonumber \\
 &+& \frac{Re[g_1(Q^2) g_2(Q^2)]}{(M+M^\prime)} \left(-\frac{1}{2} \left(M^4 M^\prime -M^3 t-M^2 M^\prime ({M^\prime}^2+s)+ 
 M t ({M^\prime}^2 +2 s+t) \right. \right. \nonumber \\
 &+& \left. \left. M^\prime ({M^\prime}^2-t) (s+t)\right) \right) \Bigg], \\
C(E_e,Q^2) &=& 2 \left[ \frac{Im[f_1(Q^2) g_2(Q^2)]}{(M+M^\prime)} \left(M^2-{M^\prime}^2+t\right) + \frac{Im[f_2(Q^2) g_2(Q^2)]}
{(M + M^\prime)^2} \left(2 M t \right) \right. \nonumber \\
 &-&  \frac{Im[g_1(Q^2) g_2(Q^2)]}{(M+M^\prime)} \left(M^2+{M^\prime}^2-2 s-t\right) \Bigg], 
\end{eqnarray}
\begin{eqnarray}
N(E_e,Q^2) &=& f_1^2(Q^2) \left(\frac{1}{2} \left(-4 m_e^2 M M^\prime+2 (M^2-s) ({M^\prime}^2-s)-t \left((M-M^\prime)^2-2 s\right)
+t^2\right)\right)  \nonumber \\
&+&  \frac{f_2^2(Q^2)}{(M+M^\prime)^2} \left(\frac{1}{2} t \left(m_e^2 \left((M-M^\prime)^2-t\right)-M^4+2 M^2 s+t 
\left((M+M^\prime)^2-2 s\right)-{M^\prime}^4+2 {M^\prime}^2 s-2 s^2\right)\right) \nonumber \\
&+&  g_1^2(Q^2) \left(\frac{1}{2} \left(4 m_e^2 M M^\prime+2 (M^2-s) ({M^\prime}^2-s)-t \left((M+M^\prime)^2-2 s\right)+t^2\right)
\right) \nonumber \\
&+& \frac{|g_2( Q^2)|^2}{(M+M^\prime)^2} \left(\frac{1}{2} t \left(m_e^2 \left((M+M^\prime)^2-t\right)-M^4+2 M^2 s+t 
\left((M-M^\prime)^2-2 s\right)-{M^\prime}^4+2 {M^\prime}^2 s-2 s^2\right)\right) \nonumber \\
&+&\frac{f_1(Q^2) f_2 (Q^2)}{(M+M^\prime)}\left(\left(m_e^2-t\right) (M+ M^\prime) \left((M - M^\prime)^2-t\right) \right) 
+ f_1(Q^2) g_1(Q^2) \left(t (M^2 +{M^\prime}^2-2 s-t) \right) \nonumber \\
&+& \frac{Re[f_1(Q^2) g_2(Q^2)]}{(M+M^\prime)} \left(t (-(M - M^\prime)) (M^2+{M^\prime}^2-2 s-t) \right) \nonumber \\
&+& \frac{f_2(Q^2) g_1(Q^2)}{(M+M^\prime)} \left(t (M+M^\prime) (M^2+{M^\prime}^2-2 s-t) \right) \nonumber \\ 
&+&  \frac{Re[f_2(Q^2) g_2(Q^2)]}{(M+M^\prime)^2} \left(t (-(M-M^\prime)) (M+M^\prime) (M^2+{M^\prime}^2-2 s-t) \right) \nonumber \\
&+&  \frac{Re[g_1(Q^2) g_2(Q^2)]}{(M+M^\prime)} \left((t-m_e^2) (M - M^\prime) ((M + M^\prime)^2-t)  \right) .
\end{eqnarray}
where, $M$ and $M^\prime$ are the masses of the initial proton and the final hadron respectively, $f_{1,2} (Q^2)$ are the vector form 
factors and $g_{1,2} (Q^2)$ are the axial vector form factors. The Mandlstam variables $s$ and $t$ are given by:

\begin{eqnarray}
 s&=& m_e^2 + M^2 + 2 M E_e, \\
 t&=& M^2 + {M^\prime}^2 - 2 M E^\prime.
\end{eqnarray}
$E_e$ and $E^\prime$ are the energies of the incoming electron and the final hadron respectively.

\end{document}